\documentclass{aa501}
\usepackage{graphics}
\usepackage{epsfig}
\usepackage{subfigure}

\begin{document}

 \title{On the gas temperature in circumstellar disks around A~stars}
 \titlerunning{Gas temperature in disks around A~stars}

 \author{Inga~Kamp \and Gerd-Jan van Zadelhoff}
 \authorrunning{I.~Kamp \& G.-J. van Zadelhoff}
 \offprints{I.~Kamp (kamp@strw.leidenuniv.nl)}
 \institute{Leiden Observatory, PO Box 9513, 2300 RA Leiden, The Netherlands}
 \date{Received December 18, 2000; accepted May 4, 2001}

 \abstract{In circumstellar disks or shells it is often assumed that gas 
and dust temperatures are equal where the latter is determined by radiative
equilibrium. This paper deals with the question whether this assumption
is applicable for tenous circumstellar disks around young A~stars.
In this paper the thin hydrostatic equilibrium models described by Kamp \&
Bertoldi (\cite{Kamp}) are combined with a detailed heating/cooling balance
for the gas. The most important heating and cooling processes are
heating through infrared pumping, heating due to the drift velocity of 
dust grains, 
and fine structure and molecular line cooling. Throughout the whole disk
gas and dust are not efficiently coupled by collisions and hence their
temperatures are quite different. Most of the gas in the disk models
considered here stays well below 300~K. In the temperature range below 
$300$~K the gas chemistry is not much affected by $T_{\rm gas}$ and 
therefore the simplifying approximation $T_{\rm gas} = T_{\rm dust}$ can 
be used for calculating the chemical structure of the disk. Nevertheless 
the gas temperature is important for the quantitative interpretation of 
observations, like fine structure and molecular lines.
 \keywords{Molecular processes -- circumstellar matter -- Stars: early-type --
           Stars: individual: Vega -- Stars: individual: $\beta$~Pictoris -- 
           planetary systems}
 }

 \maketitle

 \section{Introduction}
 \label{introduct}

In the literature the expression ``Vega-type''~stars is widely used
to characterize main-sequence stars of any spectral type which show an 
infrared excess. In order to deal with a more homogeneous subgroup this 
paper concentrates on nearby A~dwarfs which are surrounded 
by dust disks including the most prominent and well studied stars 
$\beta$~Pictoris and Vega.

The literature presents a large variety of models for these 
stars ranging from simple spherical shells with constant density 
(Chini et al. \cite{Chini1}) to detailed disk models with power law 
density distributions and several dust components (Pantin et al. 
\cite{Pantin}). These models are used to derive constraints on the 
properties of the dust components like grain sizes, composition and dust 
mass in these disks.

Several attempts to detect CO or other molecules at radio wavelength 
in the disks around A~stars failed (Yamashita et al. \cite{Yamashita}; 
Dent et al. \cite{Dent}; Liseau \cite{Liseau}). On the other hand
circumstellar gas is observed in the visible spectra of several
A-type dwarfs (e.g.\ Holweger \& Rentzsch-Holm \cite{Holweger1}; 
Jolly et al. \cite{Jolly}; Welsh et al. \cite{Welsh}; Holweger et al. 
\cite{Holweger2}; Roberge et al. \cite{Roberge}) proving without 
doubt the existence of atomic and molecular gas ($\beta$~Pictoris: CO) 
in their surrounding.

In order to study the nature of the circumstellar surroundings and
the ongoing physical processes in more detail Kamp \& Bertoldi 
(\cite{Kamp}) developed disk models comprising the two components 
gas and dust. These models include a realistic treatment of the
UV radiation field and the chemistry by the use of a chemical 
reaction network. As a first approximation gas and dust are assumed to 
be effectively coupled by collisions and their temperature is derived 
from the radiation equilibrium of the dust. The basic result of that 
work is an explanation of the non-detection of CO radio lines without 
assuming a general gas depletion in the disk. The recent detection
of H$_2$ in the disk around $\beta$~Pictoris (Thi et al. \cite{Thi})
can be explained in the context of the above described chemical disk 
models.

This paper aims at a more realistic determination of the
gas temperature in the above cited disk models by the use of
a detailed heating/cooling balance. The assumption of $T_{\rm gas}
= T_{\rm dust}$ will be evaluated. The gas temperature
is important for the interpretation of observational data via
modelling of the line emission from the gas phase, e.g.\ CO radio lines, 
[\ion{C}{ii}] and [\ion{O}{i}] fine 
structure lines, and narrow circumstellar absorption lines on top 
of the broadened stellar profile like the ones seen in the 
\ion{Ca}{ii}~K line.

 \section{The circumstellar disk model}
 \label{model}

The disk model and the chemical network is described in detail by 
Kamp \& Bertoldi (\cite{Kamp}). We restrict ourselves here to a short 
summary.

The density distribution in the disk follows from a thin hydrostatic 
equilibrium model
\begin{equation}
        n(r,z) ~=~ n_{\rm i} ~(r/R_{\rm i})^{-2.5}~ e^{-z^2/2h^2}~.
        \label{eq:density}
\end{equation}
Here the dimensionless scaleheight $H \equiv h/r$ is assumed to be 
0.15. The inner radius of the disk $R_{\rm i}$ is fixed to 40~AU, the
outer radius $R_{\rm o}$ to 500~AU. The dust temperature follows from radiative 
equilibrium assuming large spherical black body grains of size $a$
\begin{equation}
T_{\rm dust} = 282.5~\left(L_{\ast}/L_\odot\right)^{1/5} 
                             \left(r/{\rm AU}\right)^{-2/5} 
                             \left(a/{\rm\mu m}\right)^{-1/5}~,
\end{equation}
with the stellar luminosity in units of the solar one $L_{\ast}/L_\odot$.
Such a disk model leads to a surface density which follows a $r^{-1.2}$ 
power law in reasonable agreement with the ``literature exponents'' of 
brightness profiles (Hayashi et al. \cite{Hayashi}; Dutrey et al. \cite{Dutrey}; 
Augereau et al. \cite{Augereau}) ranging from $-1$ to $-1.5$. The radiation 
field $F_\nu$ is described by an ATLAS9 photospheric model (Kurucz \cite{Kurucz}) 
for the appropriate stellar parameters.

We derive a stationary solution for the chemistry using a
chemical network, which consists of 47 atomic, ionic and molecular
species that are related through 260 gas-phase chemical and photoreactions.
A number of reactions is treated in more detail like H$_2$ and CO
photodissociation, and C ionisation. The only surface reactions
incorporated are H$_2$ formation and freezing out of CO on cold dust 
grain surfaces (see Kamp \& Bertoldi \cite{Kamp} for further
details).

  \subsection{The gas temperature}
  \label{temp}

Kamp \& Bertoldi (\cite{Kamp}) assumed that gas and
dust are effectively coupled by collisions and hence both have equal 
temperatures, determined by the radiative equilibrium of the dust. 
This assumption will be critically evaluated by implementing a 
detailed heating-cooling balance for the gas phase. The particle densities 
depend on the gas temperature and vice versa. Hence the energy 
balance of the gas and the chemical network have to be solved
simultaneously. The gas temperature is determined by a detailed 
energy balance $\Gamma = \Lambda$, where $\Gamma$ and $\Lambda$ are the 
sum of all relevant heating and cooling rates respectively. To find the 
solution to $\Gamma - \Lambda = 0$, we use Ridder's method, a root bracketing 
algorithm (Press et al. \cite{Press}), in a 
slightly modified way: as starting values for the gas temperature, we use
twice and half the value at the previous radial point. Each change of 
temperature due to the root bracketing is immediately followed by a 
solution of the chemical network. If the root bracketing algorithm
gets stuck, the routine starts to subsequently extend again the 
bracketing interval. In smooth areas convergence is normally achieved 
after 5 iterations.

\subsubsection{Heating processes}

In the following we give a short description of the heating processes 
taken into account for the determination of the gas temperature.

\paragraph{Photoelectric heating.}

The photoelectric emission from grain surfaces is a major heating
source for the gas (Watson \cite{Watson}; Draine \cite{Draine1}). Dust 
particles in the circumstellar disks considered here have typical 
sizes of a few microns; hence the properties of these large grains --- 
large compared to the wavelength of UV radiation --- are well described
by those of bulk material. This approach differs from the one of
Draine (\cite{Draine1}) in that we deal in this paper with micron
sized dust instead of small interstellar dust particles.
Two different grain compositions are studied: 
graphite and silicate. 

The work functions of these materials are taken from the experimental 
work of Feuerbacher \& Fitton (\cite {Feuerbacher}), $w=4.7$~eV for graphite 
and $w=8.0$~eV for silicate. Analytical fits to their data describe the 
photoelectric yields $Y$ as functions of the energy of the incoming UV 
photon $h\nu$
\begin{equation}
\log Y(h\nu) = \left\{ \begin{array}{l}
                   -19.0~\exp(-0.263\, h\nu)-1.44  \\[1mm]
                    \hspace*{4cm}\mbox{for graphite}\\[2mm]
                   -300.0~\exp(-0.577\, h\nu)-1.06 \\[1mm]
                    \hspace*{4cm}\mbox{for silicate}~.
                    \end{array}
            \right.
\end{equation}
The kinetic energy spectrum of the emitted photoelectron is crudely
approximated by Eq.(3) of Draine (\cite{Draine1})
\begin{equation}
f(E, h\nu) = (h\nu - w)^{-1}~~~~~~{\rm if}~~0 < E < h\nu - w~.
\end{equation}
Here $E$ denotes the kinetic energy of the escaping photoelectron.

The photoelectric heating rate depends on the energy of the impinging
UV photon and on the grain potential $U$. The currents that determine
the grain potential are impinging electrons and UV photons. The
contribution of impinging ions, mainly protons, is negligible, because
hydrogen is mostly neutral or even molecular in these disks.
The incident electron current is (Draine \cite{Draine1})
\begin{equation}
J_{\rm ec} = \left\{ \begin{array}{ll}
              n_{\rm e}s_{\rm e}
              \left(\displaystyle \frac{kT}{2\pi m_{\rm e}}\right)^{\frac{1}{2}}
              ~(1+\phi )        &  \hspace*{5mm}\mbox{if}~~\phi \ge 0 \\
              n_{\rm e}s_{\rm e}
              \left(\displaystyle \frac{kT}{2\pi m_{\rm e}}\right)^{\frac{1}{2}}
              ~{\rm e}^{\phi}   &  \hspace*{5mm}\mbox{if}~~\phi < 0~,
                     \end{array}
             \right.
\label{eq:jec}
\end{equation}
with $\phi\equiv eU/kT$, $m_{\rm e}$ the electron mass and $n_{\rm e}$ 
the electron density. The sticking probability of electrons to
grains $s_{\rm e}$ is
assumed to be 0.5 following the argumentation of Draine (\cite{Draine1}).

Photoelectrons will be emitted from the grains at a rate per grain
surface area given by (Draine \cite{Draine1})
\begin{equation}
J_{\rm pe} = \int_{E_{\rm min}}^{E_{\rm max}} \left(
             \int_{\nu_{\rm th}}^{\nu_{\rm max}}
              Q_{\rm abs}Y(h\nu)f(E, h\nu)F_{\nu}~d\nu \right)dE~,
\label{eq:jpe}
\end{equation}
with $E_{\rm min} = 0$ for $U < 0$, $E_{\rm min} = eU$ for $U \ge 0$
and $E_{\rm max} = h\nu_{\rm max} - w$. The absorption 
efficiency of the dust grains $Q_{\rm abs}$ is assumed to be 1 independent of 
wavelength in the UV. The frequency $\nu_{\rm th}$ denotes that of 
impinging UV photons at the threshold $w$. The equilibrium grain 
potential $U$ can be derived from equating (\ref{eq:jec}) and (\ref{eq:jpe}).

Given the grain potential and the stellar UV radiation field, the 
photoelectric heating rate can be determined from
\begin{eqnarray}
\Gamma_1 & = & 4n_{\rm H}\sigma\int_{E_{\rm min}}^{E_{\rm max}} \left(
             \int_{\nu_{\rm th}}^{\nu_{\rm max}}
              Q_{\rm abs}Y(h\nu)f(E, h\nu)F_{\nu}~d\nu \right) \nonumber \\
         &   & \hspace*{2cm}(E - eU)~dE~~~{\rm erg~cm}^{-3}~{\rm s}^{-1}~,
\end{eqnarray}
where $n_{\rm H}$ and $\sigma$ denote the particle density of hydrogen
and the grain geometric cross section per H nucleus.

In order to make the calculation of the 
photoelectric heating rate efficient
enough for the inclusion into the heating/cooling balance, the rate is 
approximated using a function similar to Bakes \& Tielens (\cite{Bakes})
\begin{equation}
\Gamma_{\rm pe} = 10^{-4}\sigma~\epsilon~\chi~n_{\rm H}
\label{eq:grpehk99}
\end{equation}
for electron particle densities $10^{-5} < n_{\rm e} < 10^5$~cm$^{-3}$, 
gas temperatures $10 < T < 10\,000$~K, and FUV photon fluxes 
$10^{-5} < \chi < 10^5$, where the flux is measured in units of the 
Habing (\cite{Habing}) field between 912--1110~\AA\
($F_{\rm H} = 1.2\,10^7$~cm$^{-2}$~s$^{-1}$).

\begin{figure}[t]
\hspace*{1mm}\epsfig{file=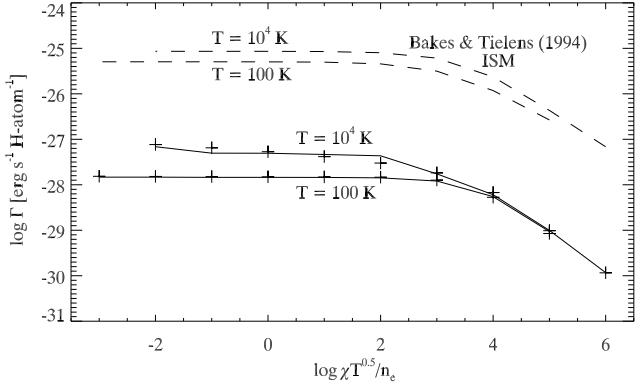, width=8.3cm}
\vspace*{-4mm}
\caption{A comparison between the exact photoelectric heating rate
(plus signs), the analytical approximation for large graphite grains
(Eq.(\ref{eq:grpehk99}): solid line, 
$\sigma=2.34\,10^{-23}$~cm$^2$~H-atom$^{-1}$) and the analytical 
approximation 
for small grains (Bakes \& Tielens \cite{Bakes}: dashed line) as a 
function of the grain charge parameter assuming ISM conditions, $\chi=1$,
and two different gas temperatures $T=100$~K and $10^4$~K.}
\label{fig:peheat_gr}
\end{figure}

\begin{figure}[h]
\hspace*{1mm}\epsfig{file=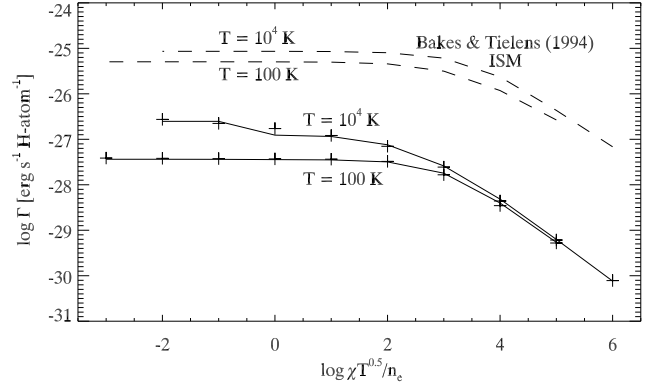, width=8.3cm}
\vspace*{-4mm}
\caption{Same as Fig.~\ref{fig:peheat_gr} but for large silicate grains}
\label{fig:peheat_si}
\end{figure}

The photoelectric efficiency $\epsilon$ 
for graphite is approximated by
\begin{equation}
\epsilon = \frac{6\,10^{-2}}{1 + 2.5\,10^{-4}~x^{0.95}}~+~
           \frac{y~(10^{-4}T)}{1 + 10^{-2}~x}~,
\label{eq:grpehapprox}
\end{equation}
where $y$ depends on the so-called grain charge parameter 
$x\equiv \chi~T^{0.5}/n_{\rm e}$
\begin{equation}
y = \left\{ \begin{array}{ll}
                   4.3\,10^{-1} & \hspace*{5mm} \mbox{if}~~~{x \leq 10^{-4}}\\[1mm]
                   2.3\,10^{-1} & \hspace*{5mm} \mbox{if}~~~
                                  {10^{-4} < x \leq 10^{-1}}\\[1mm]
                   1.5\,10^{-1} & \hspace*{5mm} \mbox{if}~~~
                                  {10^{-1} < x \leq 10^2}\\[1mm]
                   2.5\,10^{-1} & \hspace*{5mm} \mbox{if}~~~{x > 10^2}
            \end{array}
     \right.
\end{equation}
This formula yields photoelectric heating rates accurate to within 10-25\%.

The photoelectric heating rate for silicate grains is 
slightly different from the one for graphite and reads as follows
\begin{equation}
\Gamma_{\rm pe} = 2.5\,10^{-4}\sigma~\epsilon~\chi~n_{\rm H}
\label{eq:sipehk99}
\end{equation}
with
\begin{equation}
\epsilon = \frac{6\,10^{-2}}{1 + 1.8\,10^{-3}~x^{0.91}}~+~
           \frac{y~(10^{-4}T)^{1.2}}{1 + 10^{-2}~x}~,
\label{eq:sipehapprox}
\end{equation}
and
\begin{equation}
y = \left\{ \begin{array}{ll}
                   7\,10^{-1} & \hspace*{5mm} \mbox{if}~~~{x \leq 10^{-4}}\\[1mm]
                   3.6\,10^{-1} & \hspace*{5mm} \mbox{if}~~~
                                  {10^{-4} < x \leq 1}\\[1mm]
                   1.5\,10^{-1} & \hspace*{5mm} \mbox{if}~~~
                                  {x > 1}
            \end{array}
     \right.
\end{equation}

Fig.~\ref{fig:peheat_gr} compares the photoelectric heating rate derived in 
this paper for large graphite grains with that derived by Bakes \& Tielens 
(\cite{Bakes}) for ISM graphite 
\begin{equation}
\Gamma_{\rm pe} = 10^{-24}~\epsilon~\chi~n_{\rm H}~,
\label{eq:pehbt94}
\end{equation}
where the photoelectric efficiency $\epsilon$ is approximated by
\begin{equation}
\epsilon = \frac{0.0487}{1 + 4\,10^{-3}~x^{0.73}}~+~
           \frac{0.0365~(10^{-4}T)^{0.7}}{1 + 2\,10^{-4}~x}~.
\end{equation} 
The difference is due to the larger photoelectric yield of small grains. 
Photoemission is mostly a ``volume process''. In large grains the 
escape probability of photoelectrons is reduced, because they are created 
far below the surface. There scattering becomes important and they can 
easily be retrapped in the grain. Fig.~\ref{fig:peheat_si} reveals the
same effect for the large silicate grains.

\paragraph{Heating by collisional de-excitation of H$_2$.}

H$_2$ molecules excited by Lyman and Werner band absorption decay to
vib-rotationally excited levels of the ground electronic state and
heat the gas via collisional de-excitation. We 
approximate the complex process using the formula of Tielens \& 
Hollenbach (\cite{Tielens2}) for a single excited pseudovibrational level
\begin{eqnarray}
\Gamma_2 & = & f_{\rm H_2^*}~n({\rm H_2})~E_*~ 
    \left( 1.67\,10^{-11}~n({\rm H})~\sqrt{T}~{\rm e}^{-1000/T} + 
                                      \right.\nonumber \\
         &   & \left. 2.33\,10^{-11}~n({\rm H_2})~\sqrt{T}~ 
           {\rm e}^{-18100/(T+1200)} \right) \nonumber \\
         &   & {\rm erg~cm}^{-3}~{\rm s}^{-1}~.
\end{eqnarray}
$f_{\rm H_2^*}$ denotes the fixed fraction of vibrationally excited 
molecular hydrogen and $E_*=4.166\,10^{-12}\,$erg is the effective energy 
of the pseudolevel. We assume in this paper $f_{\rm H_2^*} = 10^{-5}$.

\paragraph{Heating by photodissociation of H$_2$.}

Photodissociation of H$_2$ occurs via line transitions to excited
levels followed by spontaneous radiative dissociation into two hydrogen
atoms. The kinetic energy of each H-atom is typically 0.4~eV (Stephens 
\& Dalgarno \cite{Stephens}), leading to an approximate heating rate
\begin{equation}
\Gamma_3 = 6.4\,10^{-13}~\Gamma_{\rm H_2}~n({\rm H_2})~~~
           {\rm erg~cm}^{-3}~{\rm s}^{-1}~.
\end{equation}
$\Gamma_{\rm H_2} = \chi \zeta_0 f_{\rm shield}^{\rm H_2}$~s$^{-1}$ 
denotes the H$_2$ photodissociation rate, which depends on the strength 
of the UV radiation field and on the amount of H$_2$ self-shielding 
(Kamp \& Bertoldi \cite{Kamp}).

\paragraph{Heating by H$_2$ formation on dust.}

The formation of an H$_2$ molecule on the surface of a dust grain releases
the binding energy of 4.48~eV. Due to the lack of laboratory data, we follow
the approach of Black \& Dalgarno (\cite{Black2}) and assume that this energy 
goes into translation, vibration and rotation in equal parts. Hence about 
1.5~eV will go into kinetic energy and therefore heat the gas 
\begin{equation}
\Gamma_4 = 2.39\,10^{-12}~R_{\rm form}~n({\rm H})~n_{\rm tot}~~~
           {\rm erg~cm}^{-3}~{\rm s}^{-1}~.
\end{equation}
$R_{\rm form}$ is the temperature dependent H$_2$ formation rate
(Kamp \& Bertoldi \cite{Kamp}).

\paragraph{Gas-grain collisions.}

Gas-grain collisions will act as a gas heating source for dust temperatures
larger than the gas temperature (Burke \& Hollenbach \cite{Burke}),  
\begin{eqnarray}
\Gamma_5 & = & 4.0\,10^{-12}~\pi a^2~n_{\rm tot}~n_{\rm dust}~
               \alpha_{\rm T}~\sqrt{T}~(T_{\rm dust} - T) \nonumber \\
         &   & {\rm erg~cm}^{-3}~{\rm s}^{-1}~,
\end{eqnarray}
where $n_{\rm dust}$ and $\alpha_{\rm T}$ denote the particle density of
dust grains and the thermal accomodation coefficient respectively. A typical
value for silicate and graphite dust is $\alpha_{\rm T}=0.3$ 
(Burke \& Hollenbach \cite{Burke}). If the gas temperature 
exceeds that of the dust, this rate becomes negative and therefore acts as 
a cooling rate ($\Lambda_5$).

\paragraph{Heating due to C ionisation.}

In the presence of strong UV radiation neutral carbon can be photoionized 
with an average yield of approximately 1~eV per photoelectron (Black 
\cite{Black1})
\begin{eqnarray}
\Gamma_6 & = & 1.6022\,10^{-12}~\Gamma_{\rm C}~n({\rm C})~
               f_{\rm shield}({\rm C})~\chi_{\rm x}~f_{\rm shield}({\rm H_2})
               \nonumber \\
         &   & {\rm erg~cm^{-3}~s^{-1}}~,
\end{eqnarray}
where $\chi_{\rm x}$ is the strength of the UV radiation field
attenuated by dust. The carbon self-shielding factor
\begin{equation}
f_{\rm shield}({\rm C})=\exp\left(-N({\rm C})~a_{\rm C}\right) 
\end{equation}
depends on the C column density $N({\rm C})$ and on the average carbon
ionization cross section $a_{\rm C}$. The attenuation by H$_2$ can be
approximated (Kamp \& Bertoldi \cite{Kamp}) by
\begin{equation}
f_{\rm shield}({\rm H_2})=\exp\left(-T_{\rm gas}^{0.25}~
                              \left(N({\rm H_2})/10^{22}\right)^{0.45}\right)~.
\end{equation}

\paragraph{Drift velocity of the dust grains.}

Assuming typical physical conditions for these circumstellar
disks, gas and dust are momentum coupled (Gilman \cite{Gilman}) and we
assume that all the momentum gained by the grains from the radiation
field is transferred to the gas by collisions. Dust grains of micron-size 
are usually removed by radiation pressure from the disk on timescales short 
compared to the lifetime of the disk (Artymowicz \& Clampin \cite{Artymowicz2}; 
van der Bliek et al. \cite{vanderBliek}; Backman \& Paresce \cite{Backman}; 
Artymowicz \cite{Artymowicz1}). Since observations nevertheless reveal the 
presence of micron-sized dust grains, there must be a process of continous 
replenishment for these small grains, for example destructive collisions 
between much larger dust particles.

Gas and dust in the disk rotate with slightly subkeplerian speeds 
$v_\phi$ due to the gas pressure (gas component) and due to the drag force 
between dust and gas (dust component). Hence stellar gravity is largely 
balanced by centrifugal forces. The evaluation of the drift velocities in
both directions leads to a coupled system of two-fluid two-dimensional 
hydrodynamical equations and is beyond the scope of this paper. Hence we 
restrict ourselves here to the case of maximum drift velocity, that is
assuming keplerian rotation, and bracket the ``true'' solution by 
$v_{\rm drift}=0$ and $v_{\rm drift}=v_{\rm drift}^{\rm max}$. 
The maximum drift velocity can 
then be derived from the balance of radiation pressure on the grains and drag 
force due to the grains motion through the gas (Tielens \cite{Tielens1})
\begin{equation}
v_{\rm drift}^{\rm max} = \left[ \frac{1}{2} \left( \left( 
                      \left(f_{\rm rad}
                         \right)^2 + 
                         v_{\rm gas}^4 \right)^{0.5}
                     - v_{\rm gas}^2 \right) \right]^{0.5}~{\rm cm~s}^{-1}
\label{eq:vdrift}
\end{equation}
with the thermal velocity of the gas 
\begin{equation}
v_{\rm gas} = \sqrt{\frac{5kT}{3\mu m_{\rm H}}}~~~{\rm cm~s}^{-1}
\end{equation}
and
\begin{eqnarray}
f_{\rm rad}  & = & \frac{ L_{\ast}\overline{Q}_{\rm ext} }
                               { 2\pi c\,r^2\,\rho_{\rm tot} }
               ~~~{\rm cm^2 s}^{-2}~.
\end{eqnarray}
Here $\sigma_{\rm g}$ and $m_{\rm g}$ are the geometrical cross section and
the mass of the dust grains, $L_{\ast}$ and $M_{\ast}$ are the luminosity
and mass of the central star, $\rho_{\rm tot}$ is the total mass density
of the gas, and $\overline{Q}_{\rm ext}$ is the flux-weighted
mean of the radiation pressure efficiency
\begin{equation}
\overline{Q}_{\rm ext} = \frac{\int_0^{\infty} Q_{\rm ext}(\nu)~F(\nu)~d\nu}
                              {\int_0^{\infty}F(\nu)~d\nu}~.
\end{equation}

The rate of viscous gas heating is approximated (Tielens \cite{Tielens1}) by
\begin{equation}
\Gamma_7 = 0.5~\rho_{\rm tot}~n_{\rm dust}~\sigma_{\rm g}~v_{\rm drift}^3~~~
           {\rm erg~cm}^{-3}~{\rm s}^{-1}~.
\label{eq:drift}
\end{equation}

\paragraph{Cosmic ray heating.}

Following Clavel et al. (\cite{Clavel}) The cosmic ray heating rate for
a mixture of H, H$_2$ and He is
\begin{eqnarray}
\Gamma_8 & = & (1+x_{\rm He})~\zeta({\rm H})~n({\rm H})~(1.28\,10^{-11} + 
                 2.44\,10^{-11}\,x_{\rm H_2}) \nonumber \\
         &   & {\rm erg~cm}^{-3}~{\rm s}^{-1}~,
\end{eqnarray} 
where $\zeta({\rm H}) = 6\,10^{-18}$~s$^{-1}$ is the primary cosmic ray 
ionisation
rate of hydrogen and $x_{\rm He}$ and $x_{\rm H_2}$ are the abundances
of He and molecular hydrogen relative to atomic hydrogen.

%

%

\subsubsection{Cooling processes}

In the following we give a short description of the cooling processes 
taken into account for the determination of the gas temperature.

\paragraph{\ion{O}{i} cooling.}

At densities larger than $n_{\rm cr}^{\rm H} = 
8.5\,10^5~(T/100)^{-0.69}$~cm$^{-3}$, LTE is a good approximation for 
the oxygen level occupation numbers. However the densities in the 
outer parts of these disks will be below that critical value. Therefore 
we have to calculate the statistical equilibrium (SE) of oxygen 
in detail. More importantly thermal emission of 50~K dust peaks at 60~$\mu$m, 
where we also find the [\ion{O}{i}] 63.2~$\mu$m fine-structure 
line. Since typical dust temperatures in our disks range from 20 to 150~K,
this can lead to a strong pumping of the fine-structure levels of neutral oxygen. 
An additional component to the infrared radiation field $P_\nu$
is the cosmic microwave background, which is more important for low
rotational levels of molecules like CO.

We use an oxygen model atom consisting of the lowest three fine-structure 
levels and we take into account the three fine-structure lines at 63.2, 
145.6, and 44.0~$\mu$m, and collisions with H$_2$, H, and electrons 
(Jaquet et al. \cite{Jaquet}; Launay \& Roueff \cite{Launay}; Bell et al.
\cite{Bell}). We include spontaneous emission as well as absorption and 
stimulated emission due to the IR radiation field. The importance
of the latter two processes for the statistical equilibrium calculations
is stressed by Kamp \& van Zadelhoff (\cite{Kamp2}). Table~\ref{omodellines} 
gives a short overview of the atomic data. The last column gives fits to
the critical densities above which collisional de-excitation becomes important
\begin{equation}
n_{\rm cr} = \frac{A_{ul}}{\sum_i K_{ui}}~~~{\rm cm^{-3}}~,
\end{equation}
where $K_{ui} = \epsilon_{\rm H} C_{ui}^{\rm H} + \epsilon_{\rm H_2} C_{ui}^{\rm H_2} + 
\epsilon_{\rm e^-} C_{ui}^{\rm e^-}$ denotes the sum of all collisional transitions 
from level $u$ to all other levels $i$. The critical densities tabulated refer 
to a neutral molecular gas with $\epsilon_{\rm H} = 10^{-3}$, 
$\epsilon_{\rm H_2} = 1.0$, and $\epsilon_{\rm e^-} = 10^{-6}$.

\begin{figure}[h]
\hspace*{-2mm}\epsfig{file=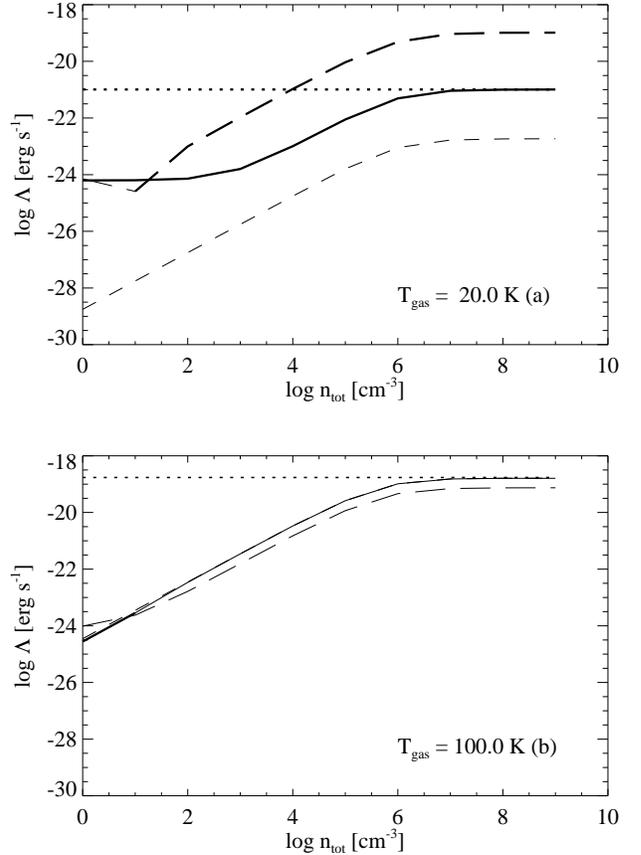, width=9cm}
\caption{[\ion{O}{i}] fine structure cooling rates for 
         (a) $T_{\rm gas}=20$~K and
         (b) $T_{\rm gas}=100$~K; solid line: SE including an IR 
         radiation field $P_\nu = 0.01\,B_\nu(80~{\rm K})
         + B_\nu(2.7~{\rm K})$, long dashed line: SE including an IR 
         radiation field $P_\nu = B_\nu(80~{\rm K})
         + B_\nu(2.7~{\rm K})$, dashed line: SE with $P_\nu = 
         B_\nu(2.7~{\rm K})$,
         dotted line: LTE including $P_\nu = 0.01\,B_\nu(80~{\rm K})
         + B_\nu(2.7~{\rm K})$; thick lines indicate a negative
         cooling rate (net heating of the gas)}
\label{OIcool}
\end{figure}

\begin{figure*}[t]
\hspace*{-1mm}\epsfig{file=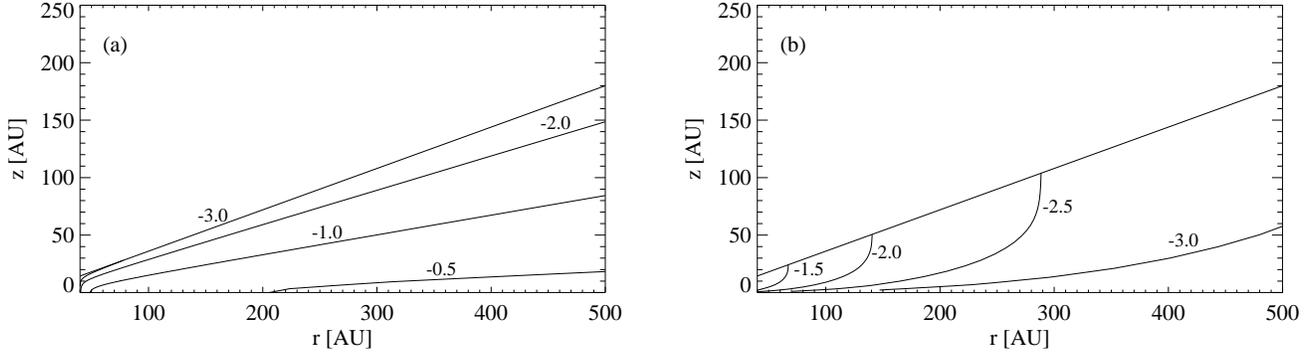, width=18cm}
\caption{Optical depth $\log \tau_{63}$ of the [\ion{O}{i}] 63.2~$\mu$m line 
         in a 2~M$_\oplus$ $\beta$~Pictoris 
         disk model:
         (a) optical depth $\log \tau$ for a photon emitted from
         the stellar surface, (b) optical depth $\log \tau$ for a
         photon emitted from the disk-midplane}
\label{O63mumdepth}
\end{figure*}

The total cooling rate is the sum over all three fine-structure transitions
\begin{eqnarray}
\Lambda_1 = \sum_k & h\nu_{ul} \biggl[ & n_u({\rm O})~
               ( A_{ul} + B_{ul}\,P_{\nu} )\,\, - \nonumber \\
                   &                  & n_l({\rm O})~B_{lu}~P_{\nu} \biggr]~~~
                       {\rm erg~cm}^{-3}~{\rm s}^{-1}~,
\label{OI63mum}
\end{eqnarray}
where $A_{ul}$ is the Einstein probability for spontaneous emission,  
$B_{ul}=c^2/(2h\nu_{ul}^3)\,A_{ul}$ is the Einstein probability for 
stimulated emission, and $B_{lu}=g_u/g_l\,B_{ul}$ the 
Einstein coefficient for absorption. The level population numbers
$n_l({\rm O})$ and $n_u({\rm O})$ refer to the lower and upper fine-structure 
level respectively, and $h\nu_{ul}$ is the energy absorbed from the total 
infrared radiation field $P_\nu$ or radiated away per emission. 
$\Lambda_1$ turns into a heating rate as soon as is becomes negative.

\begin{table}[h]
\caption{Line data for the oxygen model atom: lower and upper levels of 
         fine-structure line, statistical weights, wavelengths, transition 
         probabilities for spontaneous emission, and fits to the
         critical densities for the temperature range 
         $10~{\rm K}<T_{\rm gas}<10^3$~K} 
\begin{tabular}{cccccccc}
 u & \hspace*{-3mm}l &g$_u$& \hspace*{-4mm}g$_l$&       desig         & $\lambda$ [$\mu$m] & A$_{\rm ul}$ [s$^{-1}$] & 
  $n_{\rm cr}$ [cm$^{-3}$] \\[1mm]
\hline\\[-2mm]
 1 & \hspace*{-3mm}0 &  3  & \hspace*{-4mm} 5  & $^3$P$_1$-$^3$P$_2$ &      63.2          & $8.87\,10^{-5}$         &  
     $3.0\,10^6 T^{-0.35}$\\
 2 & \hspace*{-3mm}1 &  1  & \hspace*{-4mm} 3  & $^3$P$_0$-$^3$P$_1$ &     145.5          & $1.77\,10^{-5}$         &  
     $3.7\,10^5 T^{-0.35}$\\
 2 & \hspace*{-3mm}0 &  1  & \hspace*{-4mm} 5  & $^3$P$_0$-$^3$P$_2$ &      44.1          & $1.28\,10^{-10}$        &  
     $2.7\,T^{-0.35}$\\
\end{tabular}
\label{omodellines}
\end{table}

Fig.~\ref{OIcool} illustrates as an example the influence 
of an 80~K IR radiation field on the cooling rate. Assuming black body 
dust emission, $\Lambda_1$ is a net heating rate at low temperatures. 
Using a more ``realistic'' radiation field of $0.01\,B_\nu({\rm 80~K})$, 
which includes the dilution of radiation due to the low densities in
the disk, the heating 
rate approaches the statistical equilibrium value at particle densities 
of $10^7$~cm$^{-3}$. For higher gas temperatures ($T_{\rm gas} > 
T_{\rm dust}$) the dust 
infrared emission is no longer able to heat the gas and LTE is a
good approximation for particle densities larger than $10^7$~cm$^{-3}$.
Spontaneous emission dominates the energy balance of \ion{O}{i}
and hence the resulting cooling rate is only marginally dependent on 
the background radiation field.

The optical depth of the strongest [\ion{O}{i}] line at 63.2~$\mu$m can be 
calculated assuming that all oxygen is in the ground level, and using a 
typical line width of $\delta v_{\rm d} = 1$~km~s$^{-1}$
\begin{equation}
\tau_{10} = A_{10} \frac{\lambda_{10}^3}{8 \pi^{1.5} \delta v_{\rm d}} 
            \frac{g_1}{g_0} N_0(O)~.
\end{equation}
An optical depth $\tau_{10} = 1$ is reached for column densities of 
$N_0({\rm O}) = 3.3\,10^{17}$~cm$^{-2}$. Assuming that all oxygen is in 
atomic form and is neutral throughout the whole disk
($\log \epsilon_{\rm O} = -3.49$), this converts into a total column 
density of $N_{\rm tot} = 1.0\,10^{21}$~cm$^{-2}$. Using Eq.(4) from Kamp \&
Bertoldi (\cite{Kamp}), disks up to a total mass of 2~M$_\oplus$ are 
optically thin for [\ion{O}{i}] 63.2~$\mu$m radiation. Fig.~\ref{O63mumdepth}
illustrates the optical depth of this line in a 2~M$_\oplus$ $\beta$~Pictoris
disk model for photons emitted from the stellar surface (left) and from the 
disk-midplane (right).

\paragraph{\ion{C}{ii} cooling.}

We take the two level approximation of Hollenbach \& McKee 
(\cite{Hollenbach2}) 
for the ground state of single ionized carbon; the energies
and statistical weights are $E_0 = 0.00$~erg, $E_1 = 1.27\,10^{-14}$~erg 
and $g_0 = 2$, $g_1 = 4$ respectively. The 157.7~$\mu$m line, corresponding 
to a transition from the first excited level 2P$_{3/2}$ to the ground level 
2P$_{1/2}$, has a cooling rate of
\begin{equation}
\Lambda_2  = f_1~n({\rm \ion{C}{ii}})~A_{10}~h\nu_{10}~~~
             {\rm erg~cm}^{-3}~{\rm s}^{-1}~,
\end{equation}
where the fraction of \ion{C}{ii} in the upper level is calculated under 
the assumption of LTE
\begin{equation}
f_1 = \frac{g_1~{\rm e}^{-E_1/kT}}{g_0 + g_1~{\rm e}^{-E_1/kT}}~.
\end{equation}
The critical density for LTE $n_{\rm cr}^{\rm H} = 3.0\,10^2~
(T/100)^{-0.07}$~cm$^{-3}$ is always lower than the minimum density of 
10$^4$~cm$^{-3}$ in our disk models. The Einstein probability for 
spontaneous emission $A_{10}$ is $2.4\,10^{-6}$~s$^{-1}$.

\paragraph{H$_2$ rotational/vibrational line cooling.}

We use the H$_2$ cooling function derived by Le Bourlot, Pineau des For\^{e}ts
\& Flower (\cite{Bourlot}), which can be applied for gas temperatures
between 100 and $10^4$~K and gas densities ranging from 1 to 
$10^8$~cm$^{-3}$. Due to the wide energy spacing of the levels, rotational 
excitation of H$_2$ is negligible below 100~K. Le Bourlot, Pineau des For\^{e}ts
\& Flower take into account 51 rovibrational energy levels and their
collisional excitation by H, He and H$_2$ using recent quantum mechanical
calculations of the cross sections. Following their results the 
cooling depends only weakly
on the ortho-to-para ratio. We take $n({\rm ortho})/n({\rm para})=1.6$,
the equilibrium value at 100~K, to be representative for the low gas
temperatures in our models (see their Fig.~9).

\paragraph{\ion{O}{i} (6300~$\mu$m) cooling.}

Line emission from the metastable $^1$D level of neutral oxygen 
efficiently cools the gas at very high temperatures. We take the
cooling rate from Sternberg \& Dalgarno (\cite{Sternberg})
\begin{equation}
\Lambda_4 = 1.8\,10^{-24}~n({\rm O})~n_{\rm e}~{\rm e}^{-22\,800/T}~~~
            {\rm erg~cm}^{-3}~{\rm s}^{-1}~,
\end{equation}
where $n({\rm O})$ and $n_{\rm e}$ denote the density of neutral oxygen and 
electrons respectively.

\paragraph{Ly$\alpha$ cooling.}

The cooling by Ly$\alpha$ emission becomes efficient at very high
temperatures (Sternberg \& Dalgarno \cite{Sternberg})
\begin{equation}
\Lambda_6 = 7.3\,10^{-19}~n_{\rm e}~n({\rm H})~{\rm e}^{-118\,400/T}~~~
            {\rm erg~cm}^{-3}~{\rm s}^{-1}~.
\end{equation}
$n({\rm H})$ and $n_{\rm e}$ denote the density of neutral hydrogen and
electrons respectively.

\paragraph{CH cooling.}

Rotational cooling by CH is calculated using the optically thin
approximation of Hollenbach \& McKee (\cite{Hollenbach1})
\begin{equation}
L_{\rm rot} = \left\{ \begin{array}{l}
             \displaystyle \frac{4(kT)^{2}A_0}
                                {n_{\rm tot} E_0\,
                    \left(1\,+\,(n_{\rm cr}/n_{\rm tot})\,
                        +\,1.5(n_{\rm cr}/n_{\rm tot})^{0.5}\right)}     \\[4mm] 
 {\rm erg~s}^{-1}~{\rm cm}^3~~~~~\mbox{if}~~~n_{\rm tot} \gg n_{\rm cr} \\[6mm]
             \displaystyle \frac{kT\,(1\,-\,n({\rm H_2})/n_{\rm tot})\, 
                     \sigma_{\rm tot}v_{\rm T}}
                  {1\,+\,(n_{\rm tot}/n_{\rm cr})\, 
                       +\,1.5(n_{\rm tot}/n_{\rm cr})^{0.5}}            \\[4mm]  
 {\rm erg~s}^{-1}~{\rm cm}^3~~~~~\mbox{if}~~~n_{\rm tot} \ll n_{\rm cr}
                      \end{array} \right.
\end{equation}
where $n_{\rm cr}$, $A_0$ and $E_0$ are 
$6.6\,10^9~T_3^{-1/2}$~cm$^{-3}$, $7.7\,10^{-3}$~s$^{-1}$ and 
$2.76\,10^{-15}$~erg respectively for the CH radical. 
$v_{\rm T}$ is the thermal velocity of colliding hydrogen atoms. The 
cooling rate is then
\begin{equation}
\Lambda_7 = n_{\rm tot}~n_{\rm CH}~L_{\rm rot}~~~
                {\rm erg~cm}^{-3}~{\rm s}^{-1}~,
\end{equation}
where $n_{\rm CH}$ denotes the particle density of the CH molecule.

\paragraph{CO cooling}
\label{cocooling}

For the CO molecule the occupation numbers of the rotational lines in the
ground-vibrational state differ significantly from LTE for densities
below $10^7$~cm$^{-3}$. Therefore we use a CO model molecule to calculate 
the statistical equilibrium of carbon monoxide. It consists of 26 levels up 
to J=25 and has 351 CO-H$_2$ collisional rate coefficients (Schinke et al. 
\cite{Schinke}). The pumping of CO levels
by IR radiation is included in the same way as for \ion{O}{i}
(see respective paragraph). The molecular line data is summarized in 
Table~\ref{comodellines}. The last column of this table gives the best fits
to the critical 
densities for the CO rotational lines using the same assumptions as made
for the critical densities of \ion{O}{i}.

\begin{figure}[h]
\hspace*{-2mm}\epsfig{file=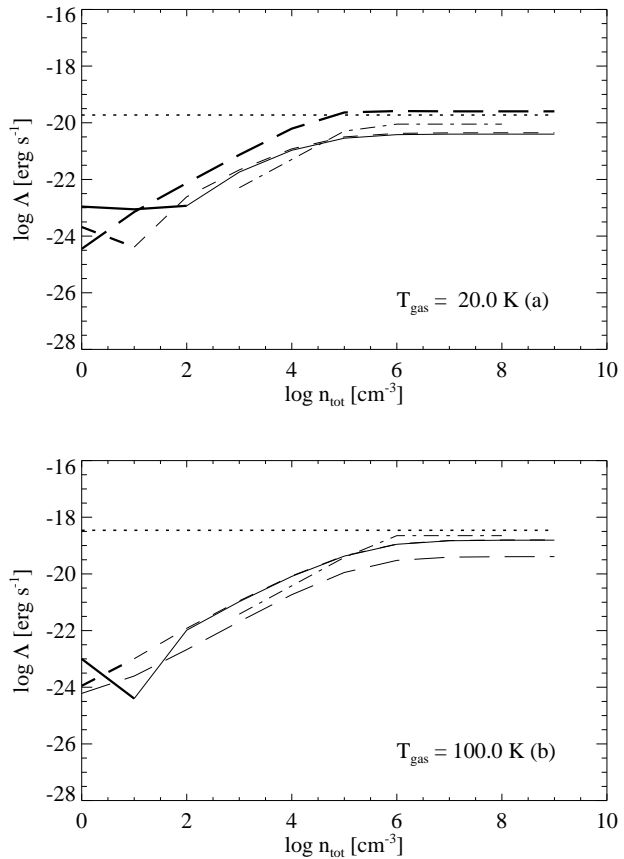, width=9cm}
\caption{CO rotational line cooling rates for (a) $T_{\rm gas}=20$~K and
         (b) $T_{\rm gas}=100$~K; solid line: SE including an IR 
         radiation field $P_\nu = 0.01\,B_\nu(80~{\rm K})
         + B_\nu(2.7~{\rm K})$, long dashed line: SE including an IR 
         radiation field $P_\nu = B_\nu(80~{\rm K})
         + B_\nu(2.7~{\rm K})$, dashed line: SE with $P_\nu = 
         B_\nu(2.7~{\rm K})$,
         dotted line: LTE including $P_\nu = 0.01\,B_\nu(80~{\rm K})
         + B_\nu(2.7~{\rm K})$, dash-dotted line: analytic
         approximation of McKee et al (\cite{McKee}); thick lines 
         indicate a negative cooling rate that means a net heating of the gas}
\label{COcool}
\end{figure}

\begin{table}
\caption{Molecular line data for the CO model molecule: upper level
         of the rotational lines $J\rightarrow J-1$, energy of the 
         upper level, wavelengths of the 
         lines, Einstein transition probabilities for spontaneous 
         emission, and fits to the critical densities 
         which hold for $10~{\rm K} < T_{\rm gas} < 100$~K; the statistical 
         weight of a level is $g = 2J + 1$}
\begin{tabular}{rrrrl}
 J & E [cm$^{-1}$] & $\lambda$ [mm] & A$_{\rm ij}$ [s$^{-1}$] & 
     $n_{\rm cr}$ [cm$^{-3}$]\\[1mm]
\hline\\[-2mm]
 1 &       3.85    &  2.601         &  $7.17\,10^{-8}$        & 
$5.0\,10^3 T^{-0.66}$ \\
 2 &      11.54    &  1.300         &  $6.87\,10^{-7}$        & 
$1.9\,10^4 T^{-0.45}$\\
 3 &      23.07    &  0.867         &  $2.48\,10^{-6}$        & 
$4.6\,10^4 T^{-0.35}$\\
 4 &      38.45    &  0.650         &  $6.09\,10^{-6}$        & 
$8.3\,10^4 T^{-0.28}$\\
 5 &      57.67    &  0.520         &  $1.22\,10^{-5}$        & 
$1.3\,10^5 T^{-0.22}$\\
 6 &      80.74    &  0.434         &  $2.13\,10^{-5}$        & 
$2.0\,10^5 T^{-0.18}$\\
 7 &     107.64    &  0.372         &  $3.40\,10^{-5}$        & 
$2.9\,10^5 T^{-0.16}$\\
 8 &     138.39    &  0.325         &  $5.11\,10^{-5}$        & 
$4.1\,10^5 T^{-0.14}$\\
 9 &     172.98    &  0.289         &  $7.29\,10^{-5}$        & 
$5.5\,10^5 T^{-0.12}$\\
10 &     211.40    &  0.260         &  $1.00\,10^{-4}$        & 
$7.2\,10^5 T^{-0.11}$\\
11 &     253.67    &  0.237         &  $1.33\,10^{-4}$        & 
$9.2\,10^5 T^{-0.10}$\\
12 &     299.77    &  0.217         &  $1.73\,10^{-4}$        & 
$1.1\,10^6 T^{-0.09}$\\
13 &     349.70    &  0.200         &  $2.19\,10^{-4}$        & 
$1.4\,10^6 T^{-0.08}$\\
14 &     403.46    &  0.186         &  $2.72\,10^{-4}$        & 
$1.7\,10^6 T^{-0.075}$\\
15 &     461.05    &  0.174         &  $3.33\,10^{-4}$        & 
$2.0\,10^6 T^{-0.07}$\\
16 &     522.48    &  0.163         &  $4.02\,10^{-4}$        & 
$2.4\,10^6 T^{-0.065}$\\
17 &     587.72    &  0.153         &  $4.80\,10^{-4}$        & 
$2.8\,10^6 T^{-0.06}$\\
18 &     656.79    &  0.145         &  $5.66\,10^{-4}$        & 
$3.2\,10^6 T^{-0.055}$\\
19 &     729.68    &  0.137         &  $6.60\,10^{-4}$        & 
$3.7\,10^6 T^{-0.05}$\\
20 &     806.38    &  0.130         &  $7.64\,10^{-4}$        & 
$4.2\,10^6 T^{-0.045}$\\
21 &     886.90    &  0.124         &  $8.77\,10^{-4}$        & 
$4.7\,10^6 T^{-0.04}$\\
22 &     971.23    &  0.119         &  $9.98\,10^{-4}$        & 
$5.3\,10^6 T^{-0.035}$\\
23 &    1059.37    &  0.113         &  $1.13\,10^{-3}$        & 
$5.9\,10^6 T^{-0.03}$\\
24 &    1151.32    &  0.109         &  $1.27\,10^{-3}$        & 
$6.5\,10^6 T^{-0.025}$\\
25 &    1247.06    &  0.104         &  $1.42\,10^{-3}$        & 
$6.8\,10^6$\\
\end{tabular}
\label{comodellines}
\end{table}

CO rotational cooling is calculated in the optically 
thin limit by 
summing over all 25 rotational line transitions
\begin{eqnarray}
\Lambda_8 = \sum_{\rm i=1}^{25} & h\nu_{ij} \biggl[ & n_{\rm i}({\rm CO}) 
                                ( A_{ij}\,+
                                  B _{ij}\,P_\nu )\, - \nonumber \\
                                &                   &   n_{\rm j}({\rm CO})
                                  B_{ji}\,P_\nu \biggr]~
                                  {\rm erg~cm}^{-3}~{\rm s}^{-1}~,
\end{eqnarray}
with $j=i-1$. $n_{\rm i}({\rm CO})$ denotes the occupation number 
for the level J and $A_{ij}$ is the spontaneous emission probability for the
rotational transition \mbox{J$\rightarrow$J-1}. $B_{ij}$ and $B_{ji}$ are the
Einstein coefficients for stimulated emission and absorption respectively 
and $h\nu_{ij}$ is the released or absorbed energy.
For photons travelling along the disk midplane, some of the CO lines will become
optically thick. Nevertheless this does not influence the cooling rate, 
as the energy is radiated isotropically in all directions and most of
it can therefore escape the flat disk without any reabsorption.   

\begin{figure*}[th]
\epsfig{file=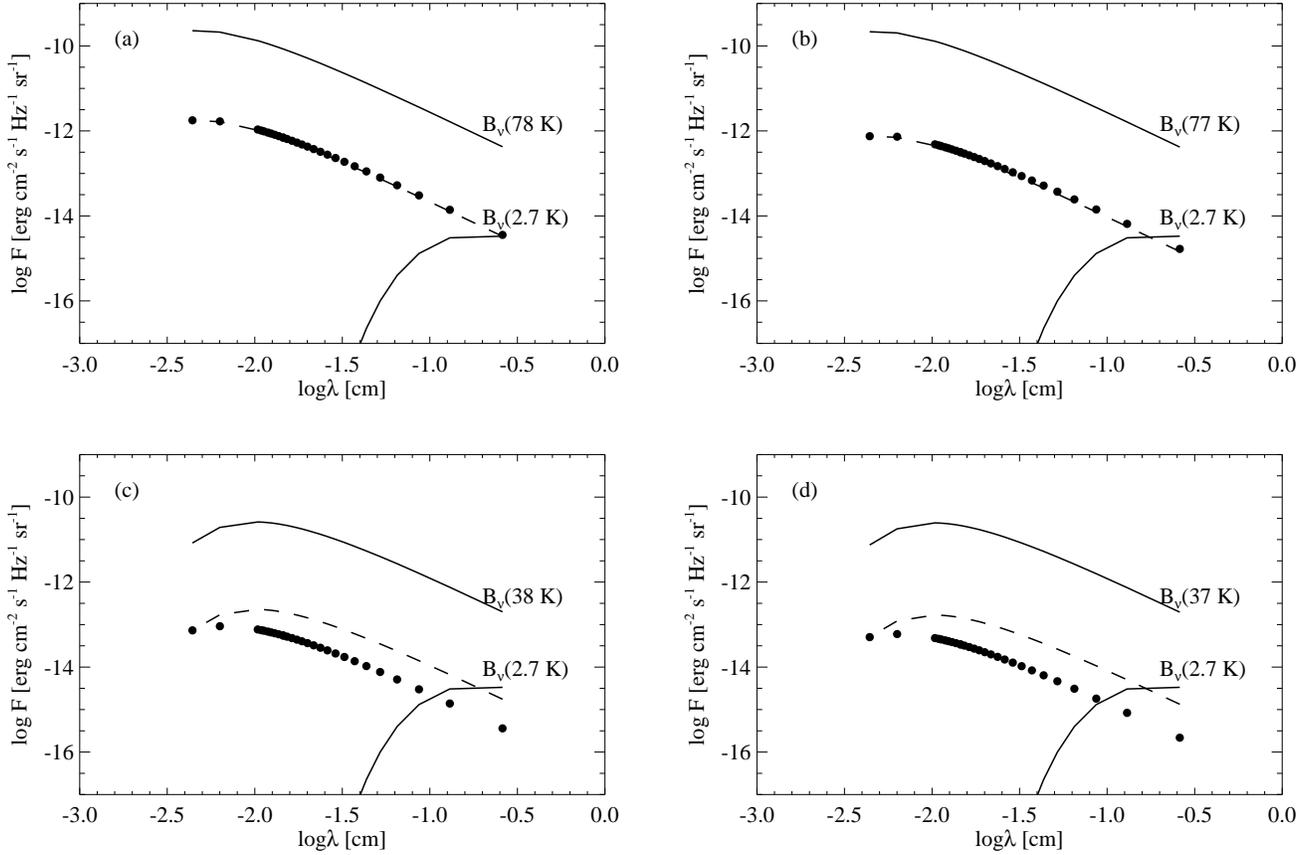, width=18cm}
\caption{Infrared radiation field for a 2~M$_\oplus$ 
         $\beta$~Pictoris disk at four different grid points:
         a) $r=50$~AU, disk midplane, b) $r=50$~AU, $z=15$~AU, c)
         $r=307$~AU, disk midplane, d) $r=307$~AU, $z=92$~AU. The black
         dots denote the radiation field due to the dust derived from 
         Eq.(\ref{Pdust}). 
         The upper solid curve shows the local approximation $P_\nu(r,z) = 
         B_\nu(T_{\rm d}(r,z))$, the lower one the cosmic microwave
         background $B_\nu(2.7~{\rm K})$. The dashed curve shows the
         local approximation scaled to the shortest wavelength point}
\label{irback}
\end{figure*}
 
McKee et al. (\cite{McKee}) determined numerical fits for the CO cooling
rate for gas temperature larger than 150~K. These cooling rates are derived
under the following assumptions: 1. optically thin lines, 2. infrared
pumping by dust emission is negligible, 3. only the ground vibrational
state contains a significant population. Assumptions 1 and 3 hold also for
our calculations, but contrary to them, we included pumping by the
infrared radiation field. Our calculations show that pumping
due to thermal dust emission of the disk and the cosmic microwave
background can be important for the CO population numbers (Kamp
\& van Zadelhoff \cite{Kamp2}). Comparing our cooling rates with 
IR pumping by the cosmic microwave background (Fig.~\ref{COcool} dashed lines)
to the numerical fits of McKee et al. (\cite{McKee}) (Fig.~\ref{COcool} 
dash-dotted lines), they agree reasonable well within a factor 2.5/3 for 
$T_{\rm gas} = 100$~K/20~K. This is within the error which the authors claim 
for their fits.

Besides this good agreement between our calculations and those of
McKee et al. (\cite{McKee}), Fig.~\ref{COcool} shows the effect
of the IR radiation field $P_\nu$ (solid line: with $P_\nu = 
0.01\,B_\nu(80~{\rm K}) + B_\nu(2.7~{\rm K})$, dashed line: with 
$P_\nu = B_\nu(2.7~{\rm K})$, long-dashed line: $P_\nu = 
B_\nu(80~{\rm K}) + B_\nu(2.7~{\rm K})$). At low particle densities
the cooling rate turns into a heating rate for the two cases
$P_\nu = 0.01\,B_\nu(80~{\rm K}) + B_\nu(2.7~{\rm K})$ and 
$P_\nu = B_\nu(2.7~{\rm K})$. Only at low temperatures and with a much
stronger radiation field $P_\nu = B_\nu(80~{\rm K}) + B_\nu(2.7~{\rm K})$
we find a net heating of the gas over the whole range of densities
considered here. For the optically thin disk models presented here, 
Fig.~\ref{COcool} and \ref{irback} reveal that the pumping of the CO 
molecule by thermal dust emission is negligible for densities larger 
than $10^3$~cm$^{-3}$ and that LTE can overestimate the cooling rates
by a factor 3 to 10 even at high densities.

\paragraph{[\ion{C}{i}] fine structure line cooling.}

Considering the chemical structure of the $\beta$~Pictoris disks (Kamp \&
Bertoldi \cite{Kamp}) [\ion{C}{i}] fine structure line cooling may 
become important
due to the presence of neutral carbon. The calculation of the cooling rates
makes use of the three level atom approximation from Hollenbach \& McKee 
(\cite{Hollenbach2}) \ion{C}{i}($^3$P$_0$, $^3$P$_1$, $^3$P$_2$). 
The respective cooling rates for the three lines at $609.2$, $229.9$, 
and $369.0$ $\mu$m are
\begin{equation}
\Lambda_{9/10/11} = A_{ij} h \nu_{ij} n_i~~~{\rm erg~cm}^{-3}~{\rm s}^{-1}~.
\end{equation}
The spontaneous transitions rates $A_{ij}$
are $7.9\,10^{-8}$, $2.0\,10^{-14}$, and $2.7\,10^{-4}$. For the 
level population numbers we can assume LTE, because the gas densities in 
these disks are well above $2\,10^3$~cm$^{-3}$, the critical density
for LTE (see Table~8 of Hollenbach \& McKee \cite{Hollenbach2}).

  \section{The infrared radiation field}

We stressed already in Sect.~\ref{temp} the importance of the infrared 
radiation field for the determination of the statistical equilibrium of 
oxygen and carbon monoxide. This radiation field consists of two components:
one is the thermal emission of the large dust grains and the other is the
2.7~K cosmic microwave background. 

Since the disks are optically thin, we can calculate the mean radiation field 
from the dust grains at any point (r,z) in the disk straight forwardly by 
integrating over the whole disk volume
\begin{eqnarray}
P_{\nu, \rm d}(r,z) & = & \frac{1}{8\pi^2} 
                            \int_0^\pi 
                            \int_{-Z_{\rm o}}^{Z_{\rm o}}
                            \int_{R_{\rm i}}^{R_{\rm o}}
                            \frac{\pi a^2}{y^2} n_{\rm d}(r',z') \nonumber \\
                    &   & \hspace*{1.5cm} B_\nu(T_{\rm d}(r',z')) 
                                        r' \,d\phi\,dz'\,dr'~,
\label{Pdust}
\end{eqnarray}
where $Z_{\rm o}$, $R_{\rm i}$ and $R_{\rm o}$ define the outer
boundary of the disk model.
The distance from $(r,z)$ to any other point $(r',z')$ in the disk is given
by
\begin{equation}
y = \sqrt{r^2 + r'^2 - 2rr'\cos \phi + (z'-z)^2}~, 
\end{equation} 
and the dust particle density can be derived from the gas density
\begin{equation}
n_{\rm d} = \delta_{\rm dg} \,n_{\rm g}\,\mu \,m_{\rm H}\, 
            \frac{3}{4\pi a^3 \rho_{\rm grain}}~~~
            {\rm cm^{-3}}~,
\end{equation}
where $\delta_{\rm dg}$ is the dust-to-gas mass ratio and 
$\rho_{\rm grain}$ is the
material density of the dust grains. The total infrared
radiation field is then the sum of the two components
\begin{equation}
P_{\nu}(r,z) = P_{\nu, \rm d}(r,z) + B_\nu(2.7~{\rm K})~.
\end{equation}

The dust particle density is too low to ensure a total $4\pi$ coverage
even in the innermost regions of the disk. For an oxygen atom sitting in
the disk plane the surrounding medium is more or less a dark background 
with a lot of bright spots, the dust grains. Hence the local approximation
$P_\nu(r,z) = B_\nu(T_{\rm d}(r,z))$ completely fails. Fig.~\ref{irback}
reveals the difference beween these two approaches for four different
points in the disk. While in the inner regions of the disk the IR 
radiation field has at least the same temperature as the local gas, this
is clearly not the case in the cool outer parts (Fig.~\ref{irback}). 
Moreover the radiation field cannot be represented by some kind 
of mean temperature scaled according to the solid angle of dust grains. 

Black body grains with a size of 3~$\mu$m have at $60~\mu$m an 
absorption efficiency of
$Q_{\rm abs} = 0.6$. The assumption of a black body radiation field is no 
longer valid when the dust temperature changes by more than 5\% over an 
optical depth of 1, which means a total column density of $7\,10^{22}$~cm$^{-2}$. 
For the disk model described here, the assumption of a black body radiation field
is valid at 50~AU for total disk masses (between 40 and 500~AU) larger than 
$7.4\,10^4$~M$_{\oplus}$ and at 100~AU for masses larger than $4.2\,10^5$~M$_{\oplus}$. 
These values are well above the disk masses considered here.

The infrared radiation field depends only on the density structure of
the disk and on the dust temperature. Since the latter is fixed by the 
assumption of radiative equilibrium, it can be calculated in advance and 
does not change during the numerical calculation of the gas chemistry and 
gas temperature.

  \section{The gas temperature: numerical models}
  \label{res:numerical}

The input parameters listed in Table~\ref{inputparameters} remain 
the same during the calculations presented here; we assume that the grains
are composed of silicates. The only parameters 
changed are the disk mass and stellar radiation field. The models
are calculated for two different disk masses, namely 2~M$_\oplus$ and
$0.2$~M$_\oplus$ and for the radiation fields of two prototype stars 
$\beta$~Pictoris and Vega, which represent the cool and the hot end in 
the range of A~stars.

\begin{table}[t]
\caption{Standard model parameters}
\begin{tabular}{lll}
    Parameter              &        Name         &  Standard value \\[1mm]
\hline\\[-2mm]
    grain size             &         $a$         & 3~$\mu$m \\
   grain density           & $\rho_{\rm grain}$  & 3~g~cm$^{-3}$ (silicates)\\
  inner radius             &     $R_{\rm i}$     & 40~AU \\
  outer radius             &     $R_{\rm o}$     & 500~AU \\
dust-to-gas mass ratio     &  $\delta_{\rm dg}$  & 0.01 \\
disk scale height          &     $H$             & 0.15 \\
fraction of H$_2^*$        &  $f_{\rm H_2^*}$    & $10^{-5}$ \\
UV dust ext. cross section &  $\sigma_{\rm UV}$  & 
                             $2.34\,10^{-23}$~cm$^2$~(H)$^{-1}$ \\    
carbon abundance           & $\epsilon_{\rm C}$  & $1.4\,10^{-4}$ \\
oxygen abundance           & $\epsilon_{\rm O}$  & $3.2\,10^{-4}$ \\
\end{tabular}
\label{inputparameters}
\end{table}

The radiation field of $\beta$~Pictoris is strong enough to blow out
3~$\mu$m silicate grains, since these disks rotate with slightly subkeplerian 
velocity around the star. The motion of the dust particles will be much
more complicated than described in Eq.(\ref{eq:vdrift}) and the velocities 
will probably be somewhat smaller. But in order to bracket reality, we have 
carried out the calculations for two extreme cases: 1. ``maximum drift velocity'', 
2. $v_{\rm drift} = 0$. We will first discuss the numerical results
including $v_{\rm drift} = v_{\rm drift}^{\rm max}$ and afterwards the results 
without drift velocity heating.

  \subsection{$\beta$~Pictoris: $v_{\rm drift} = v_{\rm drift}^{\rm max}$}

\begin{figure}[h]
\hspace*{-1mm}\epsfig{file=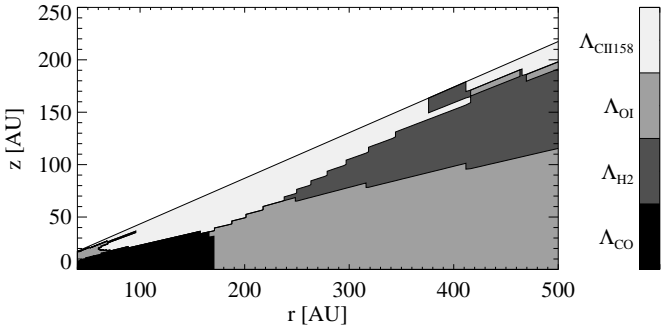, width=9cm}
\caption{The most important cooling processes in the $\beta$~Pictoris 
         2~M$_\oplus$ disk 
         model including all described heating and cooling processes
         (the bar displays only the relevant processes)}
\label{bm2.0_drift_c}
\end{figure}

\begin{figure}[h]
\hspace*{-1mm}\epsfig{file=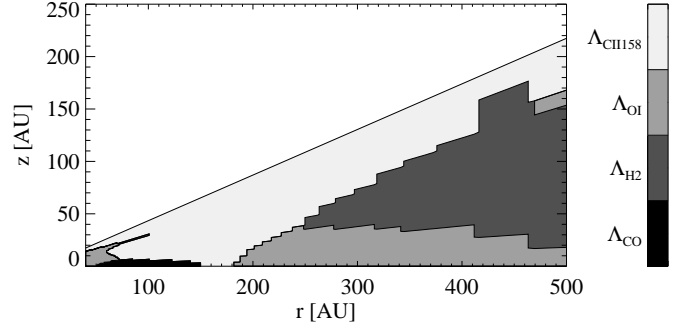, width=9cm}
\caption{The most important cooling processes in the $\beta$~Pictoris 
         $0.2$~M$_\oplus$ disk 
         model including all described heating and cooling processes
         (the bar displays only the relevant processes)}
\label{bm0.2_drift_c}
\end{figure}

Drift velocity heating dominates throughout the whole disk independent
of the disk mass in the $\beta$~Pictoris models. The most important cooling 
processes for the 2~M$_\oplus$ model are illustrated in 
Fig.~\ref{bm2.0_drift_c}. Since the stellar radiation field 
is too weak to efficiently dissociate carbon monoxide 
(Kamp \& Bertoldi \cite{Kamp}), the CO rotational lines are the most 
efficient coolant in the inner parts of the disk. Using the laboratory
data for freezing out of CO on water ice surfaces (Sandford \& Allamandola 
1990), at $T_{\rm dust} < 50$~K all CO is incorporated in ice mantles
around cold dust grains (Kamp \& Bertoldi \cite{Kamp}); hence [\ion{O}{i}] 
fine structure cooling dominates in those parts. In the uppermost
layers of the disk from inside to outside [\ion{O}{i}], [\ion{C}{ii}]
and then H$_2$ line cooling are the most important cooling processes. 
By lowering the disk mass by a factor ten (see Fig.~\ref{bm0.2_drift_c})
the dominant processes become very similar to those found in the upper layers
of the more massive disk model. The global picture does not 
change much, 
except that now [\ion{C}{ii}] and H$_2$ line cooling become more important 
compared to [\ion{O}{i}] cooling.

\begin{figure}[h]
\hspace*{-1mm}\epsfig{file=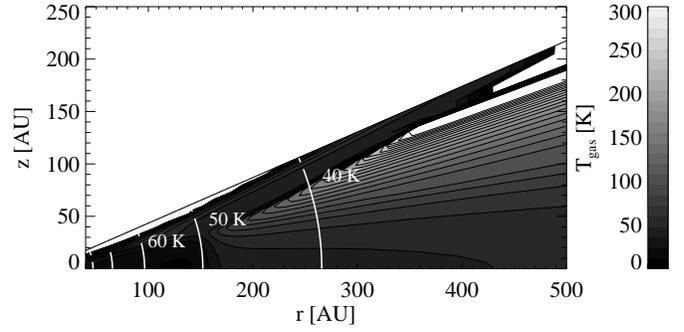, width=9cm}
\caption{Temperature structure of the $\beta$~Pictoris 2~M$_\oplus$ 
         disk model including all described heating and cooling processes: 
         the grey colors show the gas temperature as noted on the scale
         at the right hand side, while the overlayed white contour lines 
         show the dust temperature in steps of 10~K}
\label{bm2.0_drift_temp}
\end{figure}

\begin{figure}[h]
\hspace*{-1mm}\epsfig{file=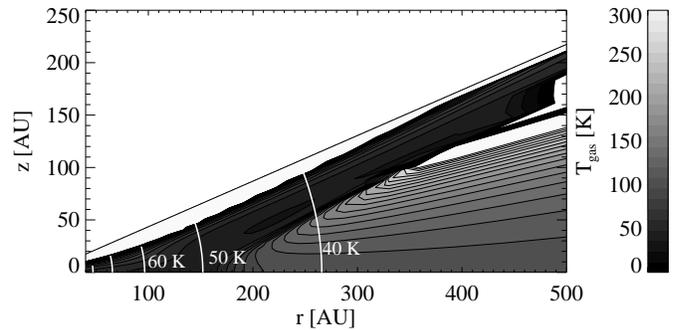, width=9cm}
\caption{Same as \ref{bm2.0_drift_temp}, but for a $\beta$~Pictoris
         $0.2$~M$_\oplus$ disk}
\label{bm0.2_drift_temp}
\end{figure}

The resulting temperature structure of both disk models is shown in
Figs.~\ref{bm2.0_drift_temp} and \ref{bm0.2_drift_temp}. Due to the 
strong drift velocity heating, the gas temperature of the disk is always 
larger than the dust temperature, except for a small core, which is below
10~K at 100~AU in the 2~M$_\oplus$ model. In both models
we find a small hot surface layer in the inner parts of the disk. We will
show later in Sect.~\ref{gaschemistry} that the total mass contained in 
this hot layer is less than 1\% of the total disk mass.

\subsection{$\beta$~Pictoris: $v_{\rm drift} = 0$}

\begin{figure}[h]
\hspace*{-1mm}\epsfig{file=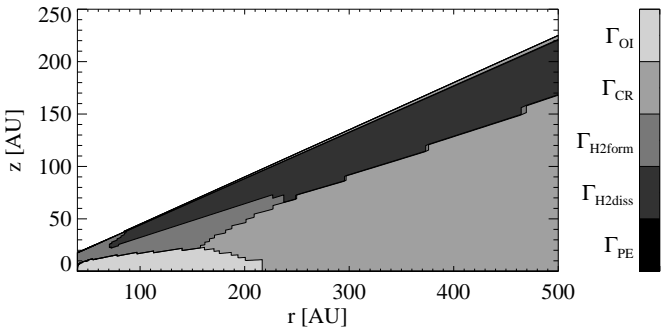, width=9cm}
\caption{The most important heating processes in the $\beta$~Pictoris 
         $2$~M$_\oplus$ disk model including all described heating and 
         cooling processes except the heating due to the drift velocity
         of grains through the gas
         (the bar displays only the relevant processes)}
\label{bm2.0_h}
\end{figure}

\begin{figure}[h]
\hspace*{-1mm}\epsfig{file=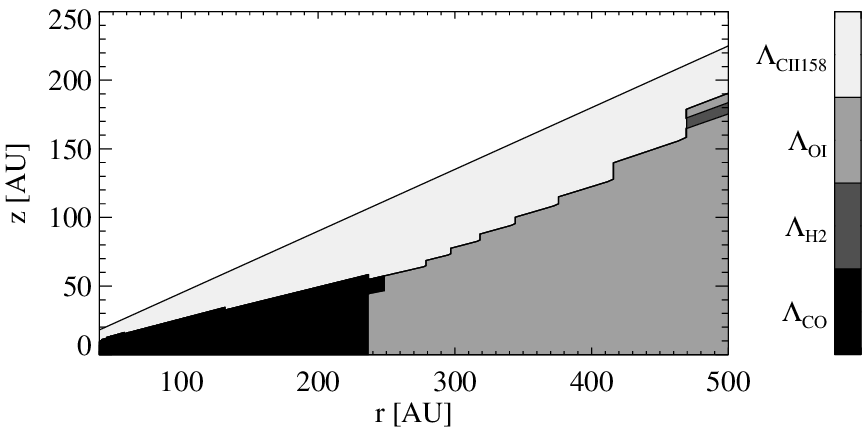, width=9cm}
\caption{The most important cooling processes in the $\beta$~Pictoris 
         $2$~M$_\oplus$ disk model including all described heating and 
         cooling processes except the heating due to the drift velocity
         of grains through the gas
         (the bar displays only the relevant processes)}
\label{bm2.0_c}
\end{figure}

If we neglect the drift velocity heating in 
the $\beta$~Pictoris disk,
cosmic ray heating becomes important in large parts of the disk (see
Fig.~\ref{bm2.0_h}). In the inner regions of the disk plane, the gas is 
heated by pumping of the \ion{O}{i} levels. Looking at the upper layers, 
we find that in the inner regions heating by H$_2$ formation dominates, 
while outwards H$_2$ dissociation is more important. On the other hand 
Fig~\ref{bm2.0_c} shows, that [\ion{C}{ii}] fine-structure cooling 
dominates in the upper disk layers, where carbon is not bound in CO. 
While in the innermost parts CO rotational line cooling contributes largely 
to the total cooling rate, more outwards, where CO is frozen onto the cool 
dust grains, [\ion{O}{i}] fine structure cooling determines the gas 
temperature.

\begin{figure}[h]
\hspace*{-1mm}\epsfig{file=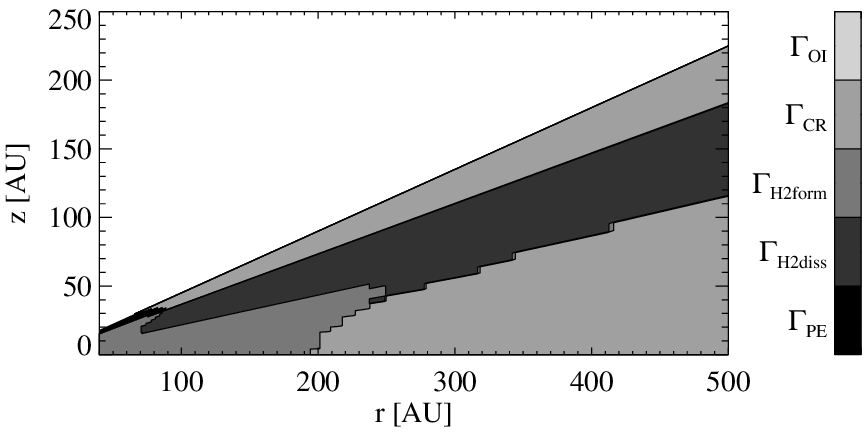, width=9cm}
\caption{The most important heating processes in the $\beta$~Pictoris 
         $0.2$~M$_\oplus$ disk model including all described heating and 
         cooling processes except the heating due to the drift velocity
         of grains through the gas
         (the bar displays only the relevant processes)}
\label{bm0.2_h}
\end{figure}

\begin{figure}[h]
\hspace*{-1mm}\epsfig{file=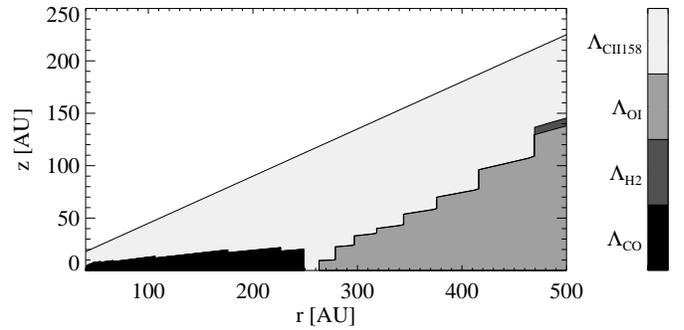, width=9cm}
\caption{The most important cooling processes in the $\beta$~Pictoris 
         $0.2$~M$_\oplus$ disk model including all described heating and 
         cooling processes except the heating due to the drift velocity
         of grains through the gas
         (the bar displays only the relevant processes)}
\label{bm0.2_c}
\end{figure}

The heating/cooling processes for the $0.2$~M$_\oplus$ model are 
shown in Figs.~\ref{bm0.2_h} and ~\ref{bm0.2_c}. Cosmic ray heating is now
also important in the uppermost low density disk layers. Between the two
regions of cosmic ray heating, the gas is heated by H$_2$ dissociation, while
in the inner region of the disk H$_2$ formation is more important. The
cooling rates show a similar behaviour as for the 2~M$_\oplus$ model except
that CO rotational line cooling and [\ion{O}{i}] fine structure cooling
are now more concentrated towards the disk midplane.

\begin{figure}[h]
\hspace*{-1mm}\epsfig{file=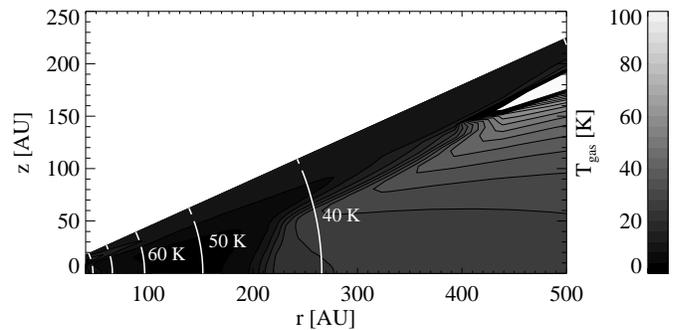, width=9cm}
\caption{Temperature structure of the $\beta$~Pictoris 2~M$_\oplus$ 
         disk model including all described heating and cooling processes
         except the heating due to the drift velocity of grains through the 
         gas: the grey colors show the gas temperature as noted on the scale
         at the right hand side, while the overlayed white contour lines 
         show the dust temperature in steps of 10~K}
\label{bm2.0_temp}
\end{figure}

\begin{figure}[h]
\hspace*{-1mm}\epsfig{file=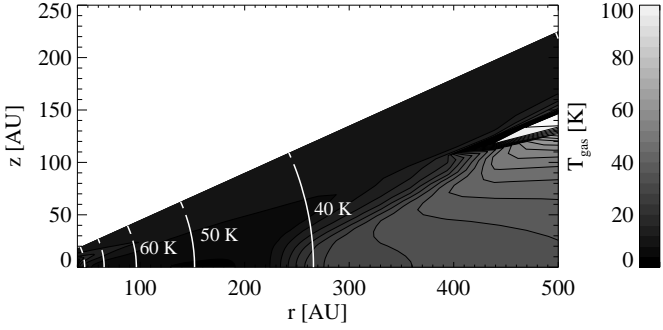, width=9cm}
\caption{Same as \ref{bm2.0_temp}, but for a $\beta$~Pictoris
         $0.2$~M$_\oplus$ disk}
\label{bm0.2_temp}
\end{figure}

Without any drift velocity heating, the gas in the disks 
becomes extremely cool, $T_{\rm gas} < 60$~K (Figs.~\ref{bm2.0_temp} and 
\ref{bm0.2_temp}).
Note that one of the main heating sources, namely heating by H$_2$ formation, 
is not as efficient as in molecular clouds, because the dust grains used in
our disks are much larger than in the ISM, and hence the dust formation
rate is lower (Kamp \& Bertoldi \cite{Kamp}).

     \subsection{Vega: $v_{\rm drift} = v_{\rm drift}^{\rm max}$}

Contrary to the $\beta$~Pictoris model, CO is mainly photodissociated
in the Vega disks and hence CO rotational line cooling is not important. 
Instead the main coolants are \ion{O}{i} and \ion{C}{ii}.

\begin{figure}[h]
\hspace*{-1mm}\epsfig{file=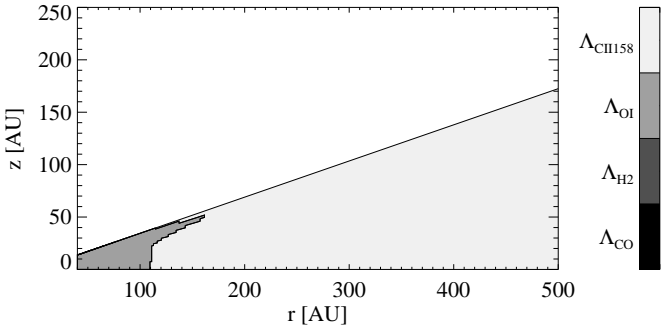, width=9cm}
\caption{The most important cooling processes in the Vega 2~M$_\oplus$ disk 
         model including all described heating and cooling processes
         (the bar displays only the relevant processes)}
\label{vm2.0_drift_c}
\end{figure}

Similar to the $\beta$~Pictoris models, drift velocity 
heating is the main energy input for the gas. Fig.~\ref{vm2.0_drift_c} 
shows, that [\ion{O}{i}] (dense inner parts) and [\ion{C}{ii}]~$157.7~\mu$m 
(outer parts) fine structure lines are the most important cooling processes 
of the 2~M$_\oplus$ Vega model.

\begin{figure}[h]
\hspace*{-1mm}\epsfig{file=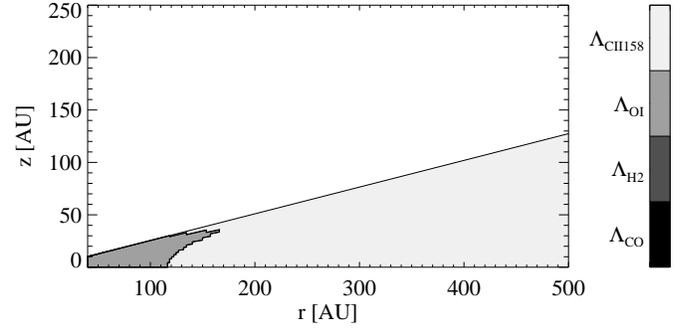, width=9cm}
\caption{The most important cooling processes in the Vega 
         $0.2$~M$_\oplus$ disk 
         model including all described heating and cooling processes
         (the bar displays only the relevant processes)}
\label{vm0.2_drift_c}
\end{figure}

If we reduce the density in the disks by a factor 10 
(Fig.~\ref{vm0.2_drift_c}), the most important cooling processes remain 
the same.

\begin{figure}[h]
\hspace*{-1mm}\epsfig{file=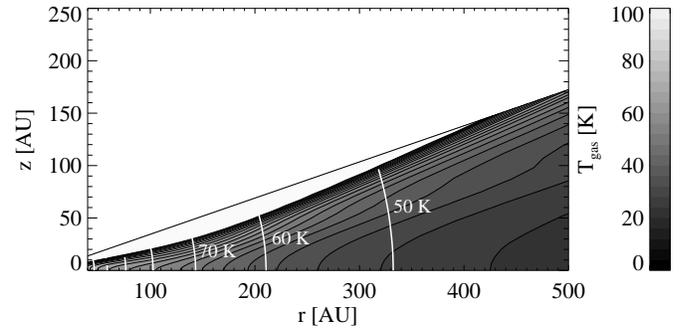, width=9cm}
\caption{Temperature structure of the Vega 2~M$_\oplus$ 
         disk model including all described heating and cooling processes:
         the grey colors show the gas temperature as noted on the scale
         at the right hand side, while the overlayed white contour lines 
         show the dust temperature in steps of 10~K}
\label{vm2.0_drift_temp}
\end{figure}

\begin{figure}[h]
\hspace*{-1mm}\epsfig{file=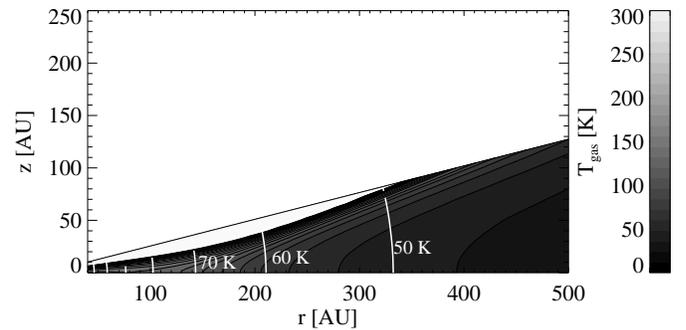, width=9cm}
\caption{Same as Fig.~\ref{vm2.0_drift_temp}, but for a $0.2$~M$_\oplus$ 
         disk}
\label{vm0.2_drift_temp}
\end{figure}

Fig.~\ref{vm2.0_drift_temp} reveals that the gas temperatures 
throughout most of the 2~M$_\oplus$ disk model stay well below 100~K. 
Nevertheless the heating due to the drift velocity of dust grains through 
the gas leads to a hot surface layer, especially in the inner
parts of the disk. Due to the lower particle densities in the 
$0.2$~M$_\oplus$ disk model this hot surface layer reaches deeper into the 
disk, leading to temperatures above 100~K even in the disk midplane. Most 
of the outer regions of the disk have nevertheless gas temperatures below 
100~K (Fig.~\ref{vm0.2_drift_temp} note the different temperature scale).

     \subsection{Vega: $v_{\rm drift} = 0$}

\begin{figure}[h]
\hspace*{-1mm}\epsfig{file=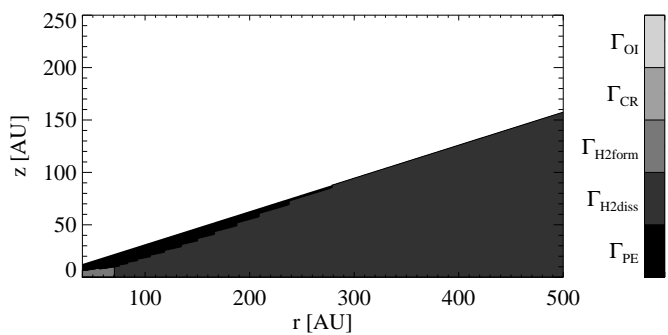, width=9cm}
\caption{The most important heating processes in the Vega 2~M$_\oplus$ disk 
         model including all described heating and cooling processes except
         the heating due to the drift velocity of grains through the 
         gas (the bar displays only the relevant processes)}
\label{vm2.0_h}
\end{figure}

\begin{figure}[h]
\hspace*{-1mm}\epsfig{file=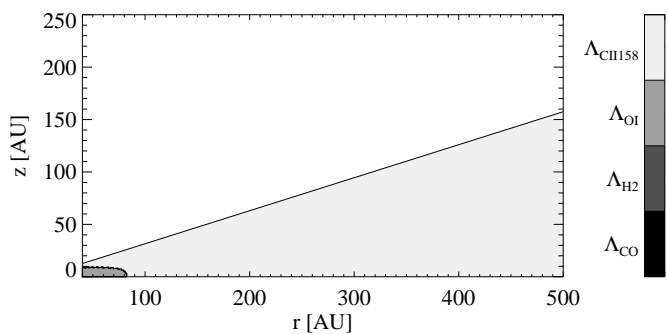, width=9cm}
\caption{The most important cooling processes in the Vega 2~M$_\oplus$ disk 
         model including all described heating and cooling processes except
         the heating due to the drift velocity of grains through the 
         gas (the bar displays only the relevant processes)}
\label{vm2.0_c}
\end{figure}

\begin{figure}[h]
\hspace*{-1mm}\epsfig{file=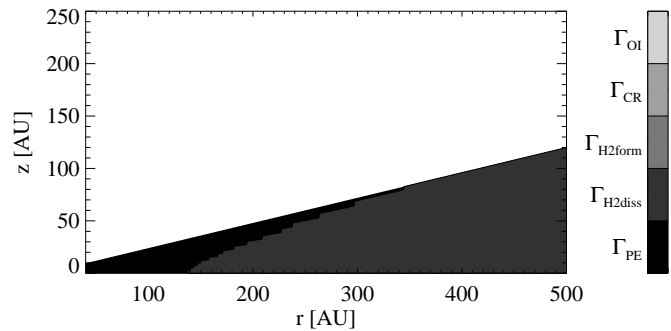, width=9cm}
\caption{The most important heating processes in the Vega $0.2$~M$_\oplus$ 
         disk 
         model including all described heating and cooling processes except
         the heating due to the drift velocity of grains through the 
         gas (the bar displays only the relevant processes)}
\label{vm0.2_h}
\end{figure}

If the drift velocity is turned off, H$_2$ dissociation and
photoelectric heating take over (Figs.~\ref{vm2.0_h} and \ref{vm0.2_h}). 
Throughout most of the 2~M$_\oplus$ disk model H$_2$ photodissociation is 
the main heating source. Only in the uppermost layers photoelectric heating
by the micron-sized grains become important. As already pointed out in 
Sect.~\ref{temp}, due to the large size of the dust grains, the photoelectric 
heating rate is 2 to 3 orders of magnitude lower than in molecular clouds with 
typical ISM grains. The main energy loss for this disk model occurs via the 
[\ion{C}{ii}]~$157.7~\mu$m fine structure line. Only in the densest parts of 
the disk midplane the [\ion{O}{i}] fine structure lines are more efficient in 
radiating away the energy (Fig.~\ref{vm2.0_c}).

\begin{figure}[h]
\hspace*{-1mm}\epsfig{file=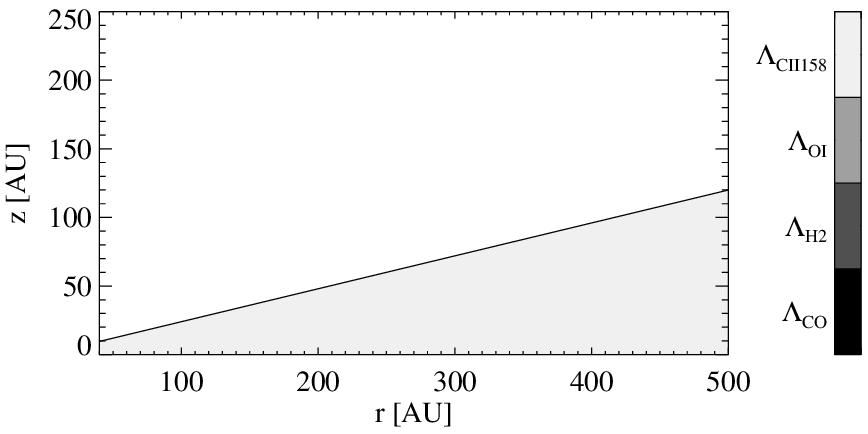, width=9cm}
\caption{The most important cooling processes in the Vega $0.2$~M$_\oplus$ 
         disk 
         model including all described heating and cooling processes except
         the heating due to the drift velocity of grains through the 
         gas (the bar displays only the relevant processes)}
\label{vm0.2_c}
\end{figure}

If we now reduce the disk mass by a factor 10, the [\ion{C}{ii}]~$157.7~\mu$m 
fine structure line is the sole remaining cooling process (see
Fig.~\ref{vm0.2_c}). On the other hand, Fig.~\ref{vm0.2_h} shows that the 
heating takes place via ejection of photoelectrons from the large dust 
grains (inner part of the modelled disk) and H$_2$ photodissociation 
(outer parts of the disk).

\begin{figure}[h]
\hspace*{-1mm}\epsfig{file=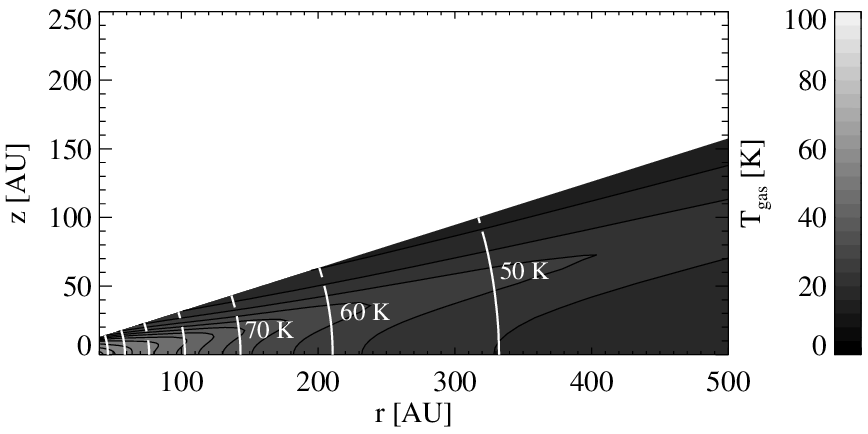, width=9cm}
\caption{Temperature structure of the Vega 2~M$_\oplus$ 
         disk model including all described heating and cooling processes
         except the heating due to the drift velocity of grains through the 
         gas: the grey colors show the gas temperature as noted on the scale
         at the right hand side, while the overlayed white contour lines 
         show the dust temperature in steps of 10~K}
\label{vm2.0_temp}
\end{figure}

\begin{figure}[h]
\hspace*{-1mm}\epsfig{file=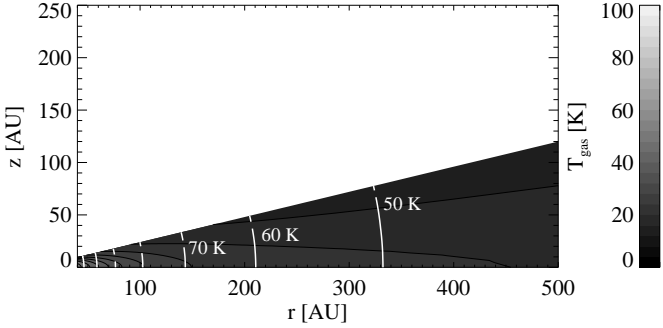, width=9cm}
\caption{Same as Fig.~\ref{vm2.0_temp}, but for the Vega $0.2$~M$_\oplus$ 
         disk model}
\label{vm0.2_temp}
\end{figure}

The resulting temperature structure of both disk models, the 2~M$_\oplus$ 
and the $0.2$~M$_\oplus$ disk model, is shown in Figs.~\ref{vm2.0_temp}
and \ref{vm0.2_temp}. The hot surface layers, which we found for the
models including the grain drift velocity, have now vanished. The whole
disk models have moderate temperatures between 10 and 100~K. Even though
the gas temperatures are in the same range as the dust temperatures, they
do not show the same structuring in the disk models. While the dust
temperature depends only on the distance from the star, the gas temperature 
depends on the particle densities and chemical structure of the disk.

  \section{The gas temperature: analytical approximations}

In the inner core of the 2~M$_\oplus$ $\beta$~Pictoris disk all carbon is 
in the form of CO, hence radiation loss by the rotational lines of CO is the 
main cooling process. On the other hand the main heating process is gas 
heating due to the drift of dust grains through the gas. 

\begin{figure}[h]
\hspace*{-1mm}\epsfig{file=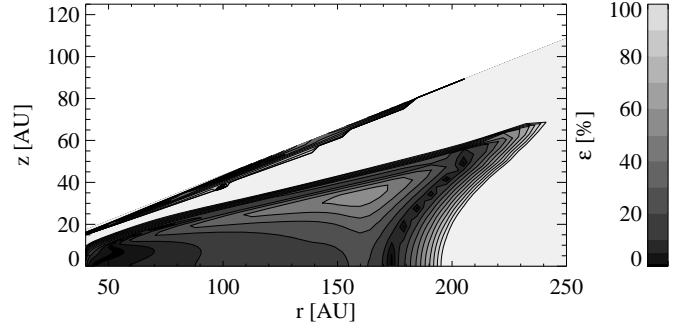, width=9cm}
\caption{Relative error of the analytical approximation for the
         gas temperature given in Eq.(\ref{tapp:1}) in the inner parts
         of the $\beta$~Pictoris 2~M$_\oplus$ model: $\epsilon = 
         (\tilde{T}_{\rm gas}-T_{\rm gas})/T_{\rm gas}$ in \%
         }
\label{tapp_bm2.0_drift}
\end{figure}

In Sect.~\ref{temp} we find that the analytical formula of McKee et al. 
(\cite{McKee}) is a good approximation for the detailed statistical
equilibrium CO cooling rate, even at low temperatures. The density 
is always larger than the critical density $n_{\rm cr}$ of McKee et 
al. (\cite{McKee})
\begin{equation}
n_{\rm cr} = 1.86\,10^4 T^{0.75}~~~{\rm cm}^{-3}~,
\end{equation}
which is a fit parameter in their formula and therefore
different from the critical densities derived in this paper for
each CO rotational line. We equate their Eq.(5.5a) modified 
by a factor of $0.5$ 
(see Fig.~\ref{COcool}) to our Eq.(\ref{eq:drift}). This gives
\begin{equation}
\tilde{T}_{\rm gas} = \sqrt{ 256.6 \left(
                    \frac{3.7\,10^{-4}}{\epsilon_{\rm CO}}\right)
                     \delta_{\rm dg} \rho_{\rm tot} 
                     \frac{v_{\rm drift}^3}{a\rho_{\rm grain}} }~~~{\rm K}~,
\label{tapp:1}
\end{equation}
where $\epsilon_{\rm CO}=n_{\rm CO}/n_{\rm tot}$ is the CO abundance, 
$\delta_{\rm dg}$ the dust-to-gas
mass ratio, $\rho_{\rm tot}$ the total gas density, and $a$ and 
$\rho_{\rm grain}$ are the grain radius and the material density of the 
dust grains respectively. Fig.~\ref{tapp_bm2.0_drift} shows the relative
error $(\tilde{T}_{\rm gas}-T_{\rm gas})/T_{\rm gas}$ of this approximation 
throughout the inner disk regions.

\begin{figure}[h]
\hspace*{-1mm}\epsfig{file=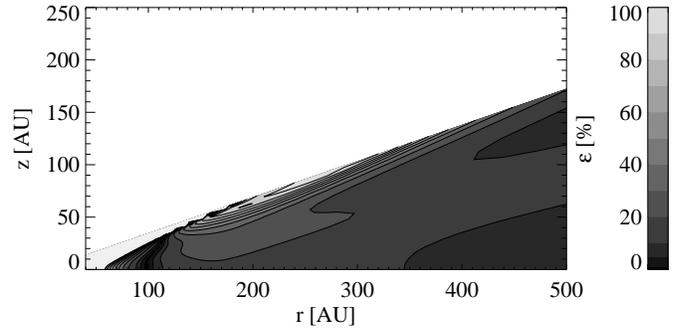, width=9cm}
\caption{Relative error of the analytical approximation for the
         gas temperature given in Eq.(\ref{tapp:2}) in the
         Vega $2.0$~M$_\oplus$ model: $\epsilon = 
         (\tilde{T}_{\rm gas}-T_{\rm gas})/T_{\rm gas}$ in \%
         }
\label{tapp_vm2.0_drift}
\end{figure}

CO is photodissociated in the Vega disks and furthermore carbon is
mainly ionized. In the presence of drift velocity heating, we can thus make 
the simplifying approximation $\Lambda_2 = \Gamma_7$, [\ion{C}{ii}] fine 
structure line cooling equals drift velocity heating for the upper disk
layers. For low gas temperatures the population number of the upper
\ion{C}{ii} level is approximately $2 e^{-91.98/T}$. This leads to the 
expression
\begin{equation}
\tilde{T}_{\rm gas} = \frac{91.98}{70.155 + 
                                  \ln(a\rho_{\rm grain}\epsilon_{\ion{C}{ii}})-
                                  \ln(n_{\rm tot}v_{\rm drift}^3)}~~~{\rm K}
\label{tapp:2}
\end{equation}
for the approximated gas temperature. Fig.~\ref{tapp_vm2.0_drift} illustrates
that this approximation can be used throughout huge parts of the Vega 
$2.0$~M$_\oplus$ disk model. Moreover it holds in the cool outer regions
of the Vega $0.2$~M$_\oplus$ disk model.

\begin{figure}[h]
\hspace*{-1mm}\epsfig{file=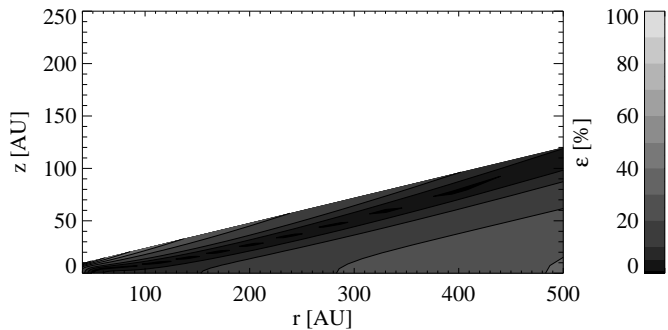, width=9cm}
\caption{Relative error of the analytical approximation for the
         gas temperature given in Eq.(\ref{tapp:3}) in the
         Vega $0.2$~M$_\oplus$ model: $\epsilon = 
         (\tilde{T}_{\rm gas}-T_{\rm gas})/T_{\rm gas}$ in \%
         }
\label{tapp_vm0.2}
\end{figure}

If we neglect drift velocity heating in these models, we can approximate
the gas temperature in the inner parts of the $0.2$~M$_\oplus$ model
by simply assuming $\Lambda_2 = \Gamma_1$, [\ion{C}{ii}] fine structure
line cooling equals photoelectric heating. For low gas 
temperatures we derive
\begin{equation}
\tilde{T}_{\rm gas} = \frac{91.98}{18.965 + \ln \epsilon_{\ion{C}{ii}}
                       - \ln \chi}~~~{\rm K}~.
\label{tapp:3}
\end{equation}
This approximation allows to calculate the disk temperature throughout
the whole Vega $0.2$~M$_\oplus$ model with an accuracy better than
about 25\% (Fig.~\ref{tapp_vm0.2}).

  \section{The gas chemistry}
\label{gaschemistry}

In order to investigate the temperature dependence of our chemical
network, we calculate the same models as discussed in 
Sect.~\ref{res:numerical} but with the assumption 
$T_{\rm gas} = T_{\rm dust}$. 

\begin{figure}[h]
\hspace*{-1mm}\epsfig{file=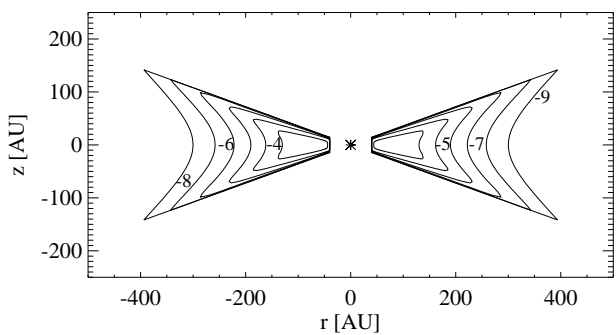, width=9cm}
\caption{$\log n_{\rm CO}/n_{\rm tot}$ in the 2~M$_\oplus$ model of
         $\beta$~Pictoris assuming $T_{\rm gas} = T_{\rm dust}$}
\label{bm2.0_td.CO}
\end{figure}

\begin{figure}[h]
\hspace*{-1mm}\epsfig{file=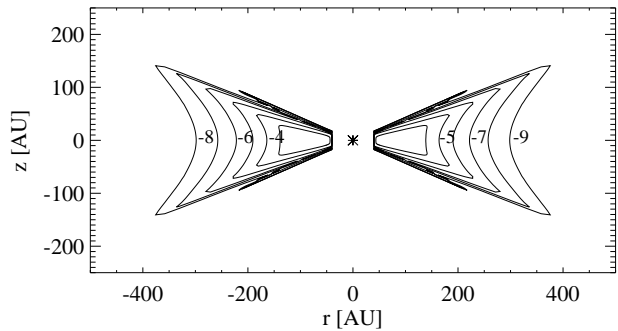, width=9cm}
\caption{$\log n_{\rm CO}/n_{\rm tot}$ in the 2~M$_\oplus$ model of
         $\beta$~Pictoris deriving $T_{\rm gas}$ from the heating/cooling
         balance including all relevant processes ($v_{\rm drift} 
         = v_{\rm drift}^{\rm max}$)}
\label{bm2.0_hc.CO}
\end{figure}

To illustrate the effect of the gas temperature on the disk chemistry, 
we show in Fig.~\ref{bm2.0_td.CO} and Fig.~\ref{bm2.0_hc.CO} the CO 
abundances in the $\beta$~Pictoris 2~M$_\oplus$ model with the 
approximation $T_{\rm gas} = T_{\rm dust}$ and with $T_{\rm gas}$ 
derived from the heating/cooling balance. The
difference in CO abundance is very small and the CO mass in
these models is $1.61\,10^{-3}$ and $1.71\,10^{-3}$~M$_\oplus$
respectively.

This shows that gas chemistry does not change very much for gas 
temperatures below 300~K. Above this threshold reactions with an activation 
barrier start to become important and they can change the overall
chemistry. Table~\ref{mass30-100K} gives the mass of material in certain
temperature ranges. The mass of material hotter than 300~K is negligible
in our disk models except in the $0.2$~M$_\oplus$ Vega model with drift
velocity heating, where it is 5\% of the total mass. On the other hand a 
large amount of the total mass can 
have gas temperatures below 30~K. Note that a low gas temperature is a 
necessary, but not sufficient criterium for freezing out of molecules. For 
this process the dust temperature plays a crucial role. The mass of material 
hotter than 100~K is the mass that can be traced by H$_2$ lines assuming the 
disks to consist of purely molecular hydrogen. This is at least for the
$\beta$~Pictoris models with the heating due to the drift velocity of
dust grains a significant percentage of the total disk mass.

\begin{table}[h]
\caption{Overview of disk masses in a certain temperature range for
         all models ($\Gamma_7$ denotes heating due to the drift velocity
         of dust grains through the gas); all masses are given in M$_\oplus$}
\begin{tabular}{llllll}
 central     &  $\Gamma_7$ &  M  &      M       &       M      &    M  \\
   star      &             &{\scriptsize total} &{\scriptsize $T_{\rm g}>300$~K}&
{\scriptsize $T_{\rm g}>100$~K}&{\scriptsize $T_{\rm g}<30$~K}         \\[1mm]
\hline\\[-2mm]
$\beta$~Pic  &   on        & 2.0 & $7\,10^{-4}$ & $0.04$       & 0.71  \\
$\beta$~Pic  &   on        & 0.2 & $10^{-3}$    & $0.10$       & $6\,10^{-5}$ \\
$\beta$~Pic  &   off       & 2.0 &  0.0         & $5\,10^{-5}$ & 1.96  \\
$\beta$~Pic  &   off       & 0.2 &  0.0         & $7\,10^{-5}$ & 0.14  \\
Vega         &   on        & 2.0 & $0.01$       & $0.05$       & 0.52  \\
Vega         &   on        & 0.2 & $0.01$       & $0.09$       & 0.00  \\
Vega         &   off       & 2.0 &  0.0         &  0.0         & 1.17  \\
Vega         &   off       & 0.2 &  0.0         &  0.0         & 0.16  \\
\end{tabular}
\label{mass30-100K}
\end{table}

  \section{Discussion}

Throughout the whole paper we discussed the two extreme cases for the
dust grain drift velocity. But is there not a better and more
accurate way to derive the drift velocity?

Klahr \& Lin (\cite{Klahr}) solved the equation of motion for
dust grains rotating in a gaseous disk around a central star.
For the implementation of the hydrodynamical drag, it is assumed
that the dust grains are in the Epstein regime, which means that
their friction time scale is much smaller than their orbital 
timescale. Unfortunately this is not fullfilled for the $3~\mu$m
grains in our disk models. Nevertheless using their Eq.(17)
for the radial drift velocity, we end up with drift velocities
which are a factor 2-5 lower than the ones we applied in our
models. Since the drift velocity enters cubic in the respective
heating term, this would lead to heating rates that are up to
a factor 125 smaller than the ones assumed in our $v_{\rm drift} =
v_{\rm drift}^{\rm max}$ models. But if we use $2~\mu$m grains 
instead of $3~\mu$m grains, we end up with drift velocities that 
are even somewhat larger than ours.

To summarize, it is obvious that the heating term due to the drift 
velocity of dust grains through the gas depends strongly on the
exact drift velocity. Unfortunately the radial drift velocity depends
also strongly on particle size, which we cannot fix exactly. Moreover 
the implementation of a more realistic hydrodynamical drag term
may change the resulting radial drift velocity considerably. In any 
case our two approaches $v_{\rm drift} = 
v_{\rm drift}^{\rm max}$ and $v_{\rm drift} = 0$ will bracket the 
real solution.

  \section{Conclusion}

The models presented in this paper are a link between the PDR models for 
dark molecular clouds (Tielens \& Hollenbach \cite{Tielens2}; Sternberg 
\& Dalgarno \cite{Sternberg}) and massive flaring disk models used for 
T Tauri stars and Herbig Ae/Be stars (Aikawa et al. \cite{Aikawa}; 
Chiang \& Goldreich \cite{Chiang}; D'Alessio et al. \cite{Alessio}).
The optically thin non-flaring disk models for young A stars presented 
by Kamp \& Bertoldi (\cite{Kamp}) are extended to self-consistently include 
the calculation of the gas temperature from a detailed heating/cooling balance.

The model calculations reveal that gas and dust temperatures in these disks
are completely different. Nevertheless there is still a coupling between 
them due to the IR pumping of \ion{O}{i} fine structure and CO rotational 
levels. Since the disks are optically thin, this coupling does not result 
in $T_{\rm gas}=T_{\rm dust}$, but instead in level population numbers that 
adjust to the local IR radiation field, which is several orders 
of magnitude below the local Planck function.

Since atomic and molecular transition probabilities are known 
with sufficient accuracy, the main uncertainty in the calculation of the gas 
temperature are the physical parameters of the dust grains, like the size and 
the composition. Except for the bright star $\beta$~Pictoris, where the 
silicate feature at $10~\mu$m is clearly detected (Telesco \& Knacke 
\cite{Telesco}; Aitken et al. \cite{Aitken}), we lack observations that 
could constrain the composition of the dust grains in these disks.

The chemical structure of these disks does not change significantly, when we
drop the approximation $T_{\rm gas}=T_{\rm dust}$ and calculate the gas
temperature from a detailed heating/cooling balance. Hence the
main results of Kamp \& Bertoldi (\cite{Kamp}) concerning the CO 
abundances in these disks do not change.

The gas temperature, the infrared radiation field and the 
statistical equilibrium calculations are crucial for the calculation of 
emission lines from these models. The results of the SE calculations
show that level population numbers can easily be orders of magnitude
different from thermodynamical equilibrium.

We will present emission line profiles and possible tracers of the gas in 
these disks in a forthcoming paper.

\paragraph{Acknowledgements}

We are indebted to F.~Bertoldi and H.~Holweger for many helpful discussions. 
Furthermore we benefited greately from discussions with E.~van~Dishoeck on the 
heating/cooling network. We thank E.~van~Dishoeck, Xander Tielens and David 
Hollenbach for a critical reading of the manuscript and helpful comments. 
This research has been supported by the ``Deutsche 
Forschungsgesellschaft'' under grant Ho 596/35-2 and by a Marie Curie 
Fellowship of the European Community programme ``Improving Human Potential'' 
under contract number MCFI-1999-00734.

\end{document}